\documentclass[aps,prc,twocolumn,superscriptaddress]{revtex4-2}

\usepackage{bm}
\usepackage{mathrsfs,amsbsy,amssymb,latexsym,amsfonts,amsmath,fancyhdr,lastpage}
\usepackage{graphicx}
\usepackage{epsfig}
\usepackage{color}
\usepackage{url}
\usepackage[normalem]{ulem}
\usepackage[hidelinks]{hyperref}
\usepackage{mhchem}
\usepackage{cases}

\definecolor{Green}{cmyk}{1,0,1,0}

\newcommand{\bbkt}[1]{\bigl\langle#1\bigr\rangle}

\newcommand{\kB}{k_{\rm B}}
\newcommand{\nm}{\nonumber\\}

\newcommand{\subA}{\mathrm{A}}
\newcommand{\subB}{\mathrm{B}}
\newcommand{\subC}{\mathrm{C}}

\newcommand{\subIN}{{n,n'-n}}

\newcommand{\nint}{n_\mathrm{m}}

\newcommand{\nA}{n_\mathrm{A}}
\newcommand{\nB}{n_\mathrm{B}}
\newcommand{\nC}{{n_\mathrm{C}}}

\newcommand{\tini}{t_\mathrm{i}}
\newcommand{\tfin}{t_\mathrm{f}}
\newcommand{\tmid}{t_\mathrm{m}}

\newcommand{\Ac}{\mathcal{A}_C}
\newcommand{\Fgain}{\Delta F^{*}}

\newcommand{\tn}{t_{\hat{n}}}

\begin{document}

\title{Gibbs Factorials Become Kinetic in History-Dependent Reactions}

\author{Nanako Hirano}
\thanks{equally contributed author}
\affiliation{Department of Physics, Ibaraki University, Mito 310-8512, Japan}

\author{Akira Yoshida}
\thanks{equally contributed author}
\email{a.yoshida.phys@gmail.com}
\affiliation{Department of Physics, Ibaraki University, Mito 310-8512, Japan}
\affiliation{Department of Physics, Kyoto University, Kyoto 606-8502, Japan}

\author{Takenobu Nakamura}
\affiliation{National Institute of Advanced Industrial Science and Technology (AIST), Tsukuba 305-8568, Japan}

\author{Naoko Nakagawa}
\email{naoko.nakagawa.phys@vc.ibaraki.ac.jp}
\affiliation{Department of Physics, Ibaraki University, Mito 310-8512, Japan}

\date{\today}

\begin{abstract}
Gibbs factorials are usually regarded as equilibrium counting factors. We show that they can also appear directly in a measurable kinetic observable when products formed at different stages are statistically distinguished in the history ensemble. In a model designed to isolate the essential ingredients, transient \ce{AB2} complexes are stabilized as \ce{C} molecules either in a single operation or through a two-stage procedure. The mean-waiting-time ratio is governed by a kinetic advantage factor $\Ac$ defined from equilibrium probabilities. A fluctuation-theorem argument identifies the dominant contribution
$\nC!/[\nint!(\nC-\nint)!]$, where \(\nint\) is the intermediate product number and \(\nC\) is the final target. This Gibbs factorial arises because products formed before and after the intermediate operation are statistically distinguished in the history ensemble, although the final molecules are macroscopically identical. Molecular dynamics simulations confirm the predicted combinatorial scaling of the mean-waiting-time ratio.
\end{abstract}

\maketitle

Gibbs factorials are usually hidden in equilibrium partition functions as counting factors for indistinguishable particles \cite{Gibbs,Pathria}. Recent studies of small thermodynamic systems have sharpened this viewpoint by showing how Gibbs factorials and mixing free energies can be assigned operational meanings \cite{Murashita-Ueda,Yoshida-Nakagawa,SHNY}. Here we ask whether such a counting factor can become kinetic.

The answer is yes when products formed at different stages are statistically distinguished in the history ensemble. When the same final products are obtained either in a single operation or through a two-stage operation separated by relaxation, the history ensemble distinguishes the products formed before and after the intermediate operation. This changes the Gibbs factorial, and hence the rare-fluctuation weight, without introducing a new product-product interaction.

This Letter formulates that effect as a kinetic statement. We introduce a kinetic advantage factor $\Ac$ from equilibrium probabilities associated with the threshold events and show that its leading contribution is fixed by a binomial Gibbs factorial. This is distinct from familiar mechanisms of enhanced reaction or cooperative response based on allosteric structural changes, conformational selection, or direct intermolecular interactions \cite{Perutz,Monod,Koshland,Cui}. Combinatorial entropy is known to influence stability, binding, and self-assembly \cite{Martinez-Frenkel,Liu2020}; the point here is that a combinatorial factor generated by product-formation history controls a measurable waiting-time ratio.

We demonstrate the mechanism in a minimal model of a reversible transient complex \ce{AB2} and an externally triggered conversion \ce{AB2 -> C}. We compare two protocols that produce the same number $\nC$ of stable \ce{C} molecules: a single conversion, in which $\nC$ transient complexes are present in the same threshold configuration, and a two-stage conversion, in which $\nint$ products are first formed and the remaining $\nC-\nint$ products are formed after relaxation. The central observable is the ratio of the mean waiting times, $\tau_1/\tau_2$.
The leading prediction is
\begin{align}
\frac{\tau_1}{\tau_2}\simeq K\frac{\nC!}{\nint!(\nC-\nint)!},
\label{e:main_kinetic_result}
\end{align}
where $K$ is a non-combinatorial kinetic prefactor.

\paragraph*{Setup: }
We consider a mixture comprising $\nA$ molecules of species \ce{A} and $\nB$ molecules of species \ce{B}. 
Hereafter, indices $i \in [1, \nA]$ and $j \in [1, \nB]$ refer to molecules of \ce{A} and \ce{B}, respectively. 
Given the phase space coordinates $\Gamma=(\Gamma_\subA,\Gamma_\subB)$ with $\Gamma_\subA\equiv(\bm{r}_{i}^\subA, \bm{p}_{i}^\subA)_{i\in[1,\nA]}$ and $\Gamma_\subB\equiv(\bm{r}_j^\subB, \bm{p}_j^\subB)_{j \in [1, \nB]}$, the classical Hamiltonian for the mixture is defined as
\begin{align}
H_0(\Gamma)=
\sum^{\nA}_{i=1} \frac{ |\bm{p}_i^\subA |^2 }{2m_\subA}+\sum^{\nB}_{j=1} \frac{ |\bm{p}_j ^\subB|^2 }{2m_\subB}
+\Psi(\{\bm{r}_i^\subA\},\{\bm{r}_j^\subB\}),
\label{e:H0}
\end{align}
where $m_\subA$ and $m_\subB$ are the masses of \ce{A} and \ce{B}. The interaction potential $\Psi(\{\bm{r}_i^{\subA}\}, \{\bm{r}_j^{\subB}\})$ is chosen so that \ce{A} and \ce{B} attract at short range, whereas identical molecules repel each other, as detailed in the Supplemental Material (SM)~\cite{SM}. These interactions transiently trap two \ce{B} molecules around one \ce{A} molecule. We adopt parameters for which this association can be viewed as the reversible reaction
\begin{align}
\ce{A + 2B <=> AB2}
\label{e:rev-reaction}
\end{align}
forming a long-lived molecular complex \ce{AB2}.

We next introduce a stable species \ce{C}. A \ce{C} molecule is formed from a transient \ce{AB2} complex by converting the dynamical binding between \ce{A} and \ce{B} into a stable binding through the irreversible operation
\begin{align}
\ce{AB2 -> C}.
\label{e:irr-reaction}
\end{align}
To describe the Hamiltonian of the mixture with $\nC$ stable \ce{C} molecules, we distinguish between atoms within \ce{C} molecules and unbound molecules.
We put superscripts `$=$' and `$\neq$' to the microstates for the molecules within \ce{C}  and unbound molecules, respectively, such as $\Gamma_\subA=(\Gamma_\subA^{=},\Gamma_\subA^{\neq})$
and $\Gamma_\subB=(\Gamma_\subB^{=},\Gamma_\subB^{\neq})$.
The total Hamiltonian is written as
\begin{align}
&H(\Gamma;\nC)
=
H_0(\Gamma_\subA^{\neq},\Gamma_\subB^{\neq})
\nm
&\qquad
+H_\subC(\Gamma_\subA^{=},\Gamma_\subB^{=};\nC)
+h_{\rm int}(\Gamma_\subA^{=},\Gamma_\subA^{\neq},\Gamma_\subB^{=},\Gamma_\subB^{\neq})
\label{e:Hamiltonian-nC}
\end{align}
with $H_0$ in \eqref{e:H0} and 
the interaction potential $h_{\mathrm{int}}$ between \ce{C} and \ce{A} or \ce{B}, which is negligible for dilute systems. $H_\subC$ is the Hamiltonian for \ce{C} molecules,
\begin{align}
H_\subC(\Gamma^{=}_\subA,\Gamma^{=}_\subB;\nC)=H_0(\Gamma^{=}_\subA,\Gamma^{=}_\subB)
+\sum_{i=1}^{\nC}\phi_{\rm b}(\bm{r}_{2i-1}^{\subB},\bm{r}_{2i}^{\subB}; \bm{r}_i^\subA),
\end{align}
where $\phi_{\rm b}$ is a binding potential that stabilizes the \ce{C} molecule, e.g.,
\begin{align}
\phi_{\rm b}(\bm{r}_{2i-1}^{\subB},\bm{r}_{2i}^{\subB}; \bm{r}_i^\subA)
=\frac{\epsilon_\mathrm{b}}{2}\sum_{j=2i-1}^{2i} (|\bm{r}^\subA_{i}-\bm{r}^{\subB}_{j}|-l)^2
\label{e:bond-potential}
\end{align}
with $l$ and $\epsilon_\mathrm{b}$ chosen such that the transient \ce{AB2} and stable \ce{C} are indistinguishable for the coarse thermodynamic observables relevant here (see~\cite{SM}). This coarse indistinguishability is essential: after conversion, the final \ce{C} molecules do not carry a visible label of their formation history, although the statistical ensemble used to describe that history does.

\paragraph*{Single- and two-stage conversion of \ce{AB2} to \ce{C}:}

The irreversible operation \eqref{e:irr-reaction} can be performed only when the transient molecular complex \ce{AB2} exists. We therefore compare two formation histories that produce the same final number $\nC$ of \ce{C} molecules: a single conversion and a two-stage conversion.

\begin{figure}[tb]
\begin{center}
\includegraphics[width=8.5cm]{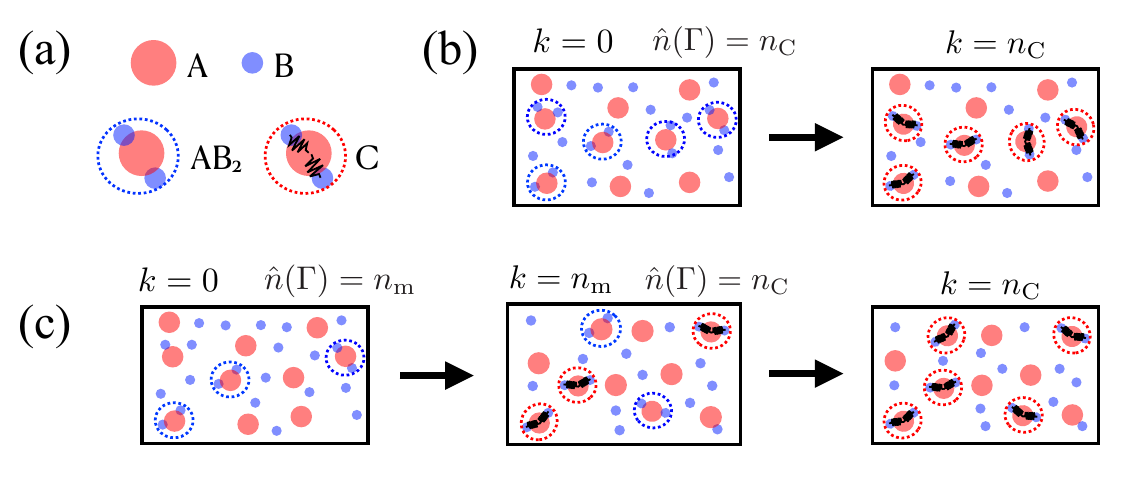} 
\end{center}
\caption{
Model and kinetic protocols. (a) Symbols for \ce{A}, \ce{B}, transient \ce{AB2}, and stable \ce{C}. (b) Single protocol: the system waits for \(\hat n(\Gamma)=\nC\) and converts the complexes to \ce{C}. (c) Staged protocol: the system first converts \(\nint\) complexes, relaxes, and then converts at \(\hat n(\Gamma)=\nC\).}
\label{fig:ndis}
\end{figure}

Because transient \ce{AB2} complexes are not stabilized by an explicit
bonding potential, we identify them using the structural criterion described
in the SM \cite{SM} and define \(\hat n(\Gamma)\) as the total number of trimers,
including both transient \ce{AB2} complexes and stable \ce{C} molecules. 
We choose the temperature and density such that
the relaxation time \(t_r\) of microscopic degrees of freedom at fixed
\(\hat n\) is shorter than the correlation time \(\tn\) of the
coarse-grained trimer-number observable used in the waiting-time measurement.
This separation supports a quasi-equilibrium description of
the threshold events used in the single and staged protocols.
Figure \ref{fig:ndis} schematically summarizes the molecular model, the definition of \(\hat{n} \), and the two formation histories compared below.

In both protocols, the system starts from an equilibrium state with no \ce{C} molecules.
In the single conversion protocol, we wait until the rare threshold \(\hat n=\nC\) is reached, i.e., $\hat n(\Gamma(\hat\tau_1))=\nC$.
We then convert the $\nC$ complexes into \ce{C} molecules at once. The mean waiting time is $\tau_1\equiv\langle\hat\tau_1\rangle$.
In the two-stage conversion protocol, we first wait until an intermediate number $\nint (< \nC)$ of complexes is formed, i.e., $\hat n(\Gamma(\hat\tau_{21}))=\nint$.
At this moment, we convert the $\nint$ complexes to \ce{C}. 
Since \(t_r<\tn\), the system relaxes in the presence of these $\nint$ stable \ce{C} molecules before the next rare event occurs.
We then wait until $\nC-\nint$ additional complexes are formed in the presence of the $\nint$ pre-existing \ce{C} molecules.
This condition is reached when $\hat n(\Gamma(\hat{\tau}_{21}+\hat{\tau}_{22}))=\nC$.
At that point, we convert these complexes to \ce{C}.
In the numerical measurement, an event is counted only when the identified transient complexes persist according to the trimer criterion, so that brief close approaches are excluded.
We define $\hat{\tau}_2\equiv \hat{\tau}_{21}+\hat{\tau}_{22}$ as the total elapsed time for the staged protocol and use the mean waiting times
$\tau_2\equiv\langle\hat\tau_2\rangle$, $\tau_{21}\equiv\langle\hat\tau_{21}\rangle$, and $\tau_{22}\equiv\langle\hat\tau_{22}\rangle$.

\begin{figure}[tb]
\centering
\includegraphics[width=8.5cm]{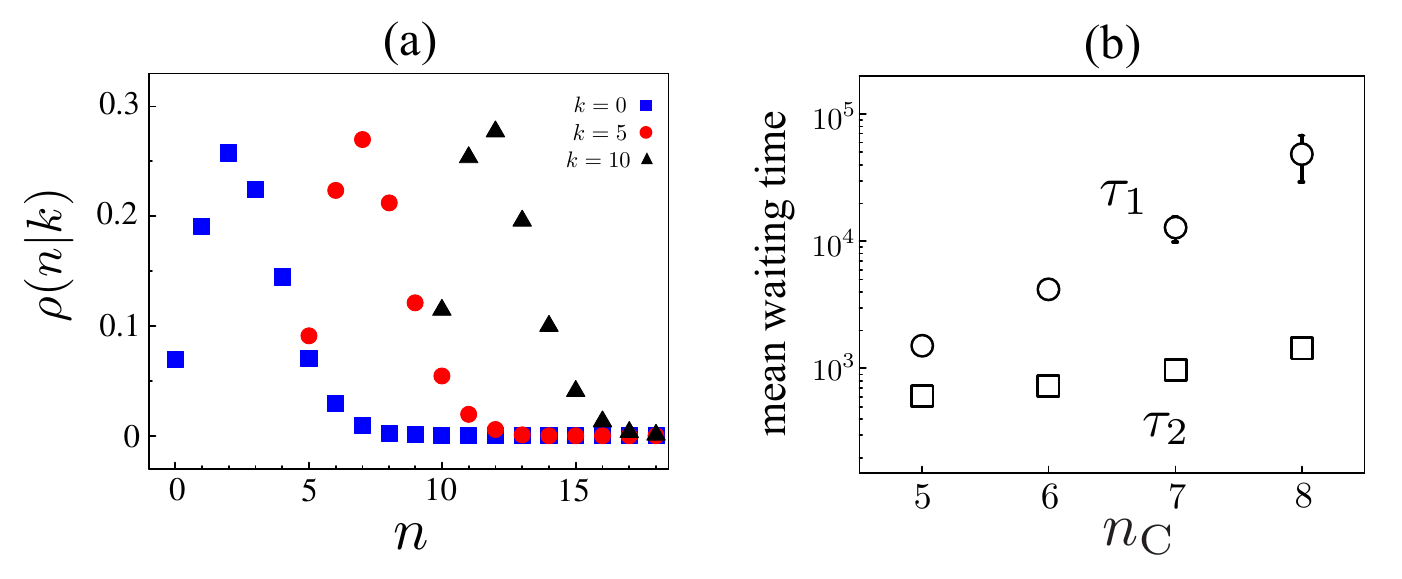} 
\caption{Equilibrium statistics and kinetic acceleration.
(a) Probabilities $\rho(n|k)$ for $k=0$, $5$, and $10$ at $\nA=108$ and $\nB=216$.
(b) Mean waiting times $\tau_1$ and $\tau_2$ for producing $\nC$ stable \ce{C} molecules, with $\nint=4$ for $\tau_2$.
For the waiting-time measurement, see Sec.~VIII of the SM~\cite{SM}.
Error bars indicate standard errors.
}
\label{fig:compare}
\end{figure}

The probability of observing $n$ trimers, given that $k$ molecules of \ce{C} are already present, is formally given by the canonical average:
\begin{align}
\rho(n|k) &= \int d\Gamma~ \delta_{n,\hat{n}(\Gamma;k)}
\frac{e^{-\beta H(\Gamma;k)} }{Z(k)}
\label{e:rho-def}
\end{align}
with
$Z(k)=\int d\Gamma e^{-\beta H(\Gamma;k)}$.
For rare threshold events, we estimate the mean waiting times from the equilibrium tail probabilities as
\begin{align}
\tau_1\simeq\frac{\tau_0}{\rho(\nC|0)},\quad
\tau_{21}\simeq\frac{\tau_0}{\rho(\nint|0)},\quad
\tau_{22}\simeq\frac{\tau_0}{\rho(\nC|\nint)}
\label{e:tau-Kac}
\end{align}
with a characteristic time scale $\tau_0$. 
The persistence filter removes microscopic recrossings and renormalizes the attempt time \(\tau_0\), while the rare-event weight in \eqref{e:tau-Kac} remains governed by the equilibrium tail \(\rho(n|k)\).
We use a common $\tau_0$ for the three threshold events.
Changing the number of pre-existing \ce{C} molecules mainly shifts the distribution of the total trimer number.
After subtracting the trivial contribution of the initial $k$ stable molecules, the similar shapes of $\rho(n|k)$ in Fig.~\ref{fig:compare}(a) suggest that the distribution of newly formed transient complexes is only weakly affected.

The expressions of the mean waiting times in \eqref{e:tau-Kac} yield 
\begin{align}
&\frac{\tau_1}{\tau_2} \simeq K \Ac(\nint, \nC),
\label{e:tau_Ac_relation}\\
&\Ac(\nint, \nC) \equiv \frac{\rho(\nC|\nint)}{\rho(\nC|0)}\frac{\rho(\nint|0)}{\rho(\nint|\nint)}.
\label{e:Ac_def}
\end{align}
Under the common-$\tau_0$ approximation, the prefactor is
\(K=\rho(\nint|\nint)/[\rho(\nint|0)+\rho(\nC|\nint)]\), the non-combinatorial contribution from the sum $\tau_2=\tau_{21}+\tau_{22}$ rather than an arbitrary fitting factor.
Equation~\eqref{e:tau_Ac_relation} motivates $\Ac$ as the kinetic advantage factor: it isolates the enhancement caused by the formation history from protocol-dependent kinetic prefactors.

\paragraph*{Kinetic Advantage as Combinatorial Factor:}

We now show that the kinetic advantage factor $\Ac$ is fixed, to leading order, by the combinatorial statistics of the formation history. Combining this result with \eqref{e:tau_Ac_relation} gives the leading prediction stated in \eqref{e:main_kinetic_result}.

To derive the factorial factor in \eqref{e:main_kinetic_result}, we formulate $\Ac$ using a fluctuation-theorem relation \cite{Jarzynski,Crooks,Seifert,SagawaUeda}.
We set the initial number of \ce{C} molecules to be \(n\) and then
perform a threshold-triggered feedback protocol to convert \(n'-n\) \ce{AB2} complexes into \ce{C} when the number of \ce{AB2} complexes is equal to \(n'-n\).
The resulting number of \ce{C} molecules is \(n'\).
In this construction, we distinguish the newly formed \(n'-n\) molecules of \ce{C} from the initial \(n\) molecules of \ce{C}. These labels specify the threshold-conditioned history ensemble. They are not molecular observables and introduce no new physical interaction.
We denote the microstate after this protocol as
\(\Gamma^\subIN=(\Gamma_\subA^{==},\Gamma_\subA^{\neq =},\Gamma_\subA^{\neq\neq},\Gamma_\subB^{==},\Gamma_\subB^{\neq =},\Gamma_\subB^{\neq\neq})\),
where
\begin{align*}
&\Gamma_\subA^{==}=(\bm{r}_{i}^\subA, \bm{p}_{i}^\subA)_{i=1}^{n},~~~~~
\Gamma_\subB^{==}=(\bm{r}_{2i-1}^{\subB}, \bm{r}_{2i}^{\subB}, \bm{p}_{2i-1}^{\subB},\bm{p}_{2i}^{\subB})_{i=1}^{n},\\
&
\Gamma_\subA^{\neq =}=(\bm{r}_{i}^\subA, \bm{p}_{i}^\subA)_{i=n+1}^{n'},~~
\Gamma_\subB^{\neq =}=(\bm{r}_{2i-1}^{\subB}, \bm{r}_{2i}^{\subB}, \bm{p}_{2i-1}^{\subB},\bm{p}_{2i}^{\subB})_{i=n+1}^{n'},\\
&\Gamma_\subA^{\neq \neq}=(\bm{r}_{i}^\subA, \bm{p}_{i}^\subA)_{i=n'+1}^{\nA},~
\Gamma_\subB^{\neq\neq}=(\bm{r}_{j}^\subB, \bm{p}_{j}^\subB)_{j=2n'+1}^{\nB}.
\end{align*}
\((\Gamma_{\subA}^{\neq =},\Gamma_{\subB}^{\neq =})\) specify the microstates for the newly formed \(n'-n\) molecules of \ce{C} and
\((\Gamma_{\subA}^{==},\Gamma_{\subB}^{==})\) are the microstates for the initial \(n\) molecules of \ce{C}.
\(\Gamma_{\subA}^{\neq \neq}\) and \(\Gamma_{\subB}^{\neq \neq}\) are the microstates for the unbound \ce{A} and \ce{B} molecules throughout the protocol.
From the Hamiltonian $H(\Gamma^\subIN;n')$, we define free energy
\begin{align}
&F(n,n'-n) \equiv -\kB T \ln\frac{Z(n')}{(\nA-n')!(n'-n)!n! (\nB-2n')!~2^{n'}}.
\label{e:F_nC_thermo}
\end{align}
$F(n,n'-n)$ is not the free energy $F(n')$ for the mixture containing $n'$ molecules of \ce{C}.
Rather, $F(n)$ satisfies $F(n,0)=F(0,n)=F(n)$.
The factorials in the denominators of \eqref{e:F_nC_thermo} take into account the indistinguishability of molecules, as specified in the expression of $\Gamma^\subIN$.
Thus the \ce{A} indices are partitioned into the initial products, the newly formed products, and the unbound molecules, with sizes \(n\), \(n'-n\), and \(\nA-n'\). The \ce{B} indices are partitioned into \(2n'\) bound atoms and \(\nB-2n'\) unbound atoms; the two \ce{B} atoms within each \ce{C} molecule are indistinguishable, giving the factor \(2^{n'}\).

Equilibrium distributions are thermostat independent; for the path-probability argument, we take Langevin dynamics, for which local detailed balance has the standard form.
Suppose the time evolution of $\Gamma$ is Markovian with the transition probability  \(\mathcal{T}_{n}(\hat{\Gamma})\) for a path \(\hat{\Gamma}\) with \(n\) molecules of \ce{C}.
Let \(\tmid\) be the time at which \(n'-n\) molecules of \ce{AB2} bind to form \ce{C}.
Precisely, the \ce{A} and \ce{B} molecules to be bound as \ce{C} at \(t=\tmid\) are not yet identified for \(t<\tmid\).
By picking out the trajectories that have \(n'-n\) complexes \ce{AB2} at \(t=\tmid\),
we denote the microstate at \(t>\tmid\) as \(\Gamma^\subIN(t)\)
and define a path \(\hat{\Gamma}_>\) for \(\tmid < t \leq \tfin\) as
\(\hat{\Gamma}_> \equiv (\Gamma^\subIN(t))_{t\in(\tmid,\tfin]} \).
We extend it to \(\tini \le t<\tmid\) as \(\hat{\Gamma}_<= (\Gamma^\subIN(t))_{t\in[\tini,\tmid)} \).
The entire path is then written as
\(\hat{\Gamma}_{n\to n'}=(\hat{\Gamma}_<, ~ \hat{\Gamma}_>)\),
and its time-reversed path as  \(\hat{\Gamma}_{n\to n'}^\dagger=(\hat{\Gamma}_>^\dagger, ~ \hat{\Gamma}_<^\dagger)\).
The local detailed balance condition is then
\begin{align}
{\mathcal{T}_{n}(\hat{\Gamma}_<)\mathcal{T}_{n'}(\hat{\Gamma}_>)}
=e^{-\beta Q}{\mathcal{T}_{n'}(\hat{\Gamma}_>^\dagger)\mathcal{T}_{n}(\hat{\Gamma}_<^\dagger)},
\label{eEM:LDB}
\end{align}
where
\begin{align}
Q=&H(\Gamma^\subIN(\tfin);n')-H(\Gamma^\subIN(\tini);n)
-W_{n,n'}.
\label{eEM:Q-def}
\end{align}
The work performed in the conversion from \(n\) to \(n'\) stable \ce{C} molecules is denoted by \(W_{n,n'}\) and given by
\begin{align}
W_{n,n'} = \frac{\epsilon_\mathrm{b}}{2}\sum_{i=n+1}^{n'} \sum_{j=2i-1}^{2i} (|\bm{r}^\subA_{i}-\bm{r}^{\subB}_{j}|-l)^2.
\label{e:work-def}
\end{align}
Multiplying both sides of \eqref{eEM:LDB} by the Kronecker delta
and applying \eqref{eEM:Q-def} leads to
\begin{align}
&e^{-\beta H(\Gamma^\subIN(\tini);n)}\mathcal{T}_{n}(\hat{\Gamma}_<)\delta_{n',\hat{n}(\Gamma^\subIN(\tmid))} 
e^{-\beta W_{n,n'}}\mathcal{T}_{n'}(\hat{\Gamma}_>)
\nm
&
=e^{-\beta H(\Gamma^\subIN(\tfin);n')}\mathcal{T}_{n'}(\hat{\Gamma}_>^\dagger)\delta_{n',\hat{n}(\Gamma^\subIN(\tmid))}
\mathcal{T}_{n}(\hat{\Gamma}_<^\dagger).
\label{eEM:LDB2}
\end{align}
Integrating \eqref{eEM:LDB2} over all paths, \(\int \mathcal{D}\hat{\Gamma}_<\mathcal{D}\hat{\Gamma}_> \),
we have
\begin{align}
e^{-\beta F(n)}\rho(n'|n)\bbkt{e^{-\beta W_{n,n'}}}_{n'}
=
e^{-\beta F(n,n'-n)}\rho(n'|n'),
\label{eEM:formula2}
\end{align}
as illustrated in Appendix \ref{sec:EM_Derivation}.
Here \(\langle\cdot\rangle_{n'}\) denotes the average conditioned on the threshold event \(\hat n(\Gamma)=n'\).

For three cases \((n,n')=(0,\nint)\),  \((\nint,\nC)\), and \((0,\nC)\),
\eqref{eEM:formula2} yields
\begin{align}
&e^{-\beta F(0)}\rho(\nint|0)\bbkt{e^{-\beta W_{0,\nint}}}_{\nint} =e^{-\beta F(0,\nint)}\rho(\nint|\nint),\nm
&e^{-\beta F(\nint)}\rho(\nC|\nint)\bbkt{e^{-\beta W_{\nint,\nC}}}_{\nC} 
=e^{-\beta F(\nint,\nC-\nint)}\rho(\nC|\nC),\nm
&e^{-\beta F(0)}\rho(\nC|0)\bbkt{e^{-\beta W_{0,\nC}}}_{\nC} =e^{-\beta F(0,\nC)}\rho(\nC|\nC),
\nonumber
\end{align}
respectively.
Note that \(F(0,n)=F(n)\).
Combining these three equations gives
\begin{align}
&\Ac(\nint, \nC)
= \frac{\nC!}{\nint! (\nC - \nint)! } R_W,
\label{e:Ac_FT_with_work}\\
&R_W\equiv
\frac{\bbkt{e^{-\beta W_{0,\nC}}}_{\nC}}
{\bbkt{e^{-\beta W_{0,\nint}}}_{\nint}
\bbkt{e^{-\beta W_{\nint,\nC}}}_{\nC}}.
\label{e:work_correction}
\end{align}
Equation~\eqref{e:Ac_FT_with_work} is exact. For weak work fluctuations, \(\ln R_W\simeq -\beta[\bbkt{W_{0,\nC}}_{\nC}-\bbkt{W_{0,\nint}}_{\nint}-\bbkt{W_{\nint,\nC}}_{\nC}]\), which is nearly vanishing and negligible in the present model. We then obtain the dominant contribution
\begin{align}
&\Ac(\nint, \nC)
\simeq \frac{\nC!}{\nint! (\nC - \nint)! },
\label{e:Ac_FT_general}
\end{align}
which yields \eqref{e:main_kinetic_result}.
Equivalently, the same result may be written as a history-dependent free-energy offset
\begin{align}
\Fgain(\nint,\nC)
&\equiv F(\nC)-F(\nint,\nC-\nint)\nm
&= \kB T \ln \frac{\nC!}{\nint!(\nC-\nint)!}.
\label{e:Fgain_definition}
\end{align}
In this Letter we use \(\Ac\) as the central object because it is defined from the equilibrium probabilities associated with the threshold events and directly controls the kinetic ratio \(\tau_1/\tau_2\); \(\Fgain\) is the thermodynamic representation of its Gibbs-factorial contribution.

\begin{figure}[bt] 
\begin{center} 
\includegraphics[width=7.5cm]{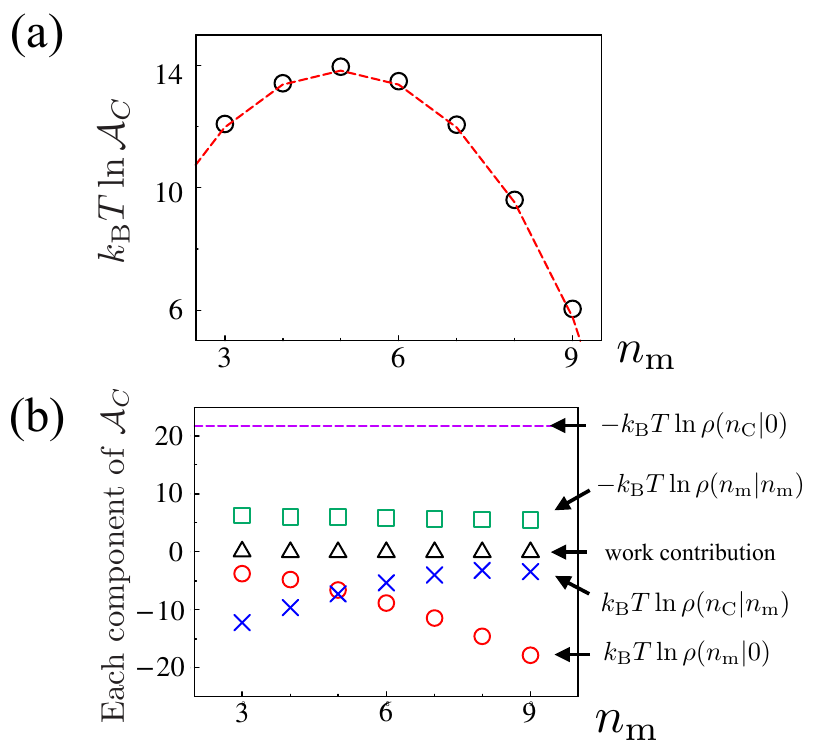} 
\end{center} 
\caption{Numerical verification of the Gibbs-factorial contribution for $\nC=10$ at $\nA=108$ and $\nB=216$. (a) $\kB T\ln \Ac$ from simulated probabilities. (b) Terms in \(\kB T\ln\Ac\), including the nearly vanishing and negligible work contribution associated with \(R_W\). Standard-error bars are smaller than the symbols.}
\label{fig:Gibbs} 
\end{figure}

\begin{figure}[bt]
\begin{center}
\includegraphics[width=5.5cm]{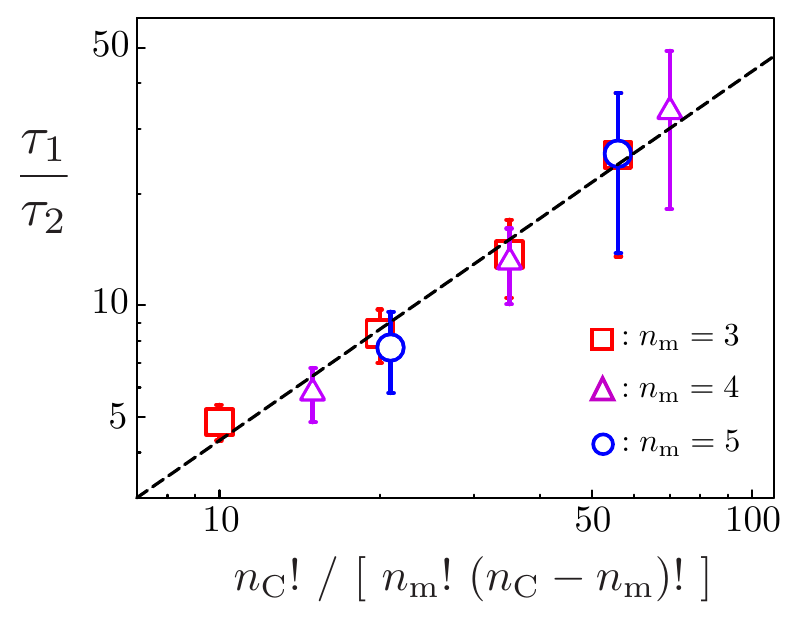} 
\end{center}
\caption{
Validation of \eqref{e:main_kinetic_result} for \(\nint=3,4,5\) and \(\nC=\nint+2,\ldots,8\).
For the waiting-time measurement, see Sec.~VIII of the SM~\cite{SM}.
The dotted line is \(K\Ac\) with \(K=0.43\).
Error bars indicate standard errors.
} 
\label{fig:Ac}
\end{figure}

\paragraph*{Numerical verification:} 
We test the prediction using molecular dynamics simulations of a dilute, low-temperature system.
Model details and parameters are given in the SM~\cite{SM}.
For the equilibrium probabilities, \(\hat n(\Gamma;k)\) is evaluated using the instantaneous geometrical criterion described in the SM.
For waiting times, we use the time-averaged-distance version of the same criterion to filter out brief close approaches without persistent complex formation.

With these preparations, we first examine the equilibrium statistics of the trimers.
Figure~\ref{fig:compare}(a) displays $\rho(n|k)$ for observing a total of $n$ trimers 
for $k=0, 5, 10$ at $\nA=108$ and $\nB=216$. As $k$ increases, the peak of $\rho(n|k)$ shifts to higher $n$. 
Figure \ref{fig:compare}(b) compares the mean waiting times for producing $\nC$ molecules of \ce{C}. $\tau_1$ is for the single conversion and $\tau_2$ is for the two-stage conversion with the intermediate step $\nint=4$.
As expected, $\tau_{1}\gg\tau_{2}$.

From the numerically determined \(\rho(n|k)\), Fig.~\ref{fig:Gibbs}(a) shows that \(\kB T\ln\Ac\) agrees with \(\Fgain\), while Fig.~\ref{fig:Gibbs}(b) shows that the work correction associated with \(R_W\) in \eqref{e:Ac_FT_with_work} is nearly vanishing and negligible.

Figure~\ref{fig:Ac} tests the kinetic consequence of this Gibbs-factorial term by comparing \(\tau_1/\tau_2\) with \(\Ac\) for several values of \(\nint\) and \(\nC\).
Over the tested range, the data are described by an empirical common prefactor \(K=0.43\) within standard errors.
Thus the residual kinetic factor \(K\) does not mask the Gibbs-factorial scaling in the present numerical tests.
A first-principles prediction of \(K\) would require a more detailed kinetic theory beyond the scope of this Letter.

\paragraph*{Discussion:}
The experimental signature is not faster staged production alone but collapse, for different histories, of \(\tau_1/\tau_2\) against the Gibbs-factorial factor in \eqref{e:main_kinetic_result}. Candidate platforms include programmable assemblies with externally stabilizable transient complexes, such as DNA-origami, DNA-coated colloids, patchy colloids, and field-controlled colloidal assemblies \cite{Dunn2015Nature,Wang2012,DiMichele2013,Gong2017,Swan2014}. Controlled stabilization triggered by a transient-complex fluctuation would provide the required operation.

\paragraph*{Acknowledgement} 

The authors thank T. Kanazawa, N. Matsubayasi, and Y. Nakayama for useful discussions and S.-i. Sasa for critical reading of the manuscript. 
The numerical simulations were performed with LAMMPS on the facilities of the Supercomputer Center, the Institute for Solid State Physics, the University of Tokyo. 
This study was supported by JSPS KAKENHI Grant Numbers JP23K22415, JP25K00923, JP25K22002, and JP26H00383.

\twocolumngrid
\appendix
\normalsize

\renewcommand{\theequation}{\Alph{section}\arabic{equation}}
\renewcommand{\thefigure}{\Alph{section}\arabic{figure}}
\renewcommand{\thetable}{\Alph{section}\arabic{table}}
\renewcommand{\theHequation}{\Alph{section}\arabic{equation}}
\renewcommand{\theHfigure}{\Alph{section}\arabic{figure}}
\renewcommand{\theHtable}{\Alph{section}\arabic{table}}

\setcounter{equation}{0}
\setcounter{figure}{0}
\setcounter{table}{0}

\vspace{1cm}

\begin{center}
{\large \bf End Matter}
\end{center}

\section{Derivation of Eq.~(\ref{eEM:formula2})}
\label{sec:EM_Derivation}

Below, we abbreviate $\Gamma^{n,n'-n}$ as $\Gamma$ for visibility.
The path integral on the right-hand side of \eqref{eEM:LDB2} yields
\begin{align}
&\int \mathcal{D}\hat{\Gamma}_>^\dagger\int \mathcal{D}\hat{\Gamma}_<^\dagger~
e^{-\beta H(\Gamma(\tfin);n')}\mathcal{T}_{n'}(\hat{\Gamma}_>^\dagger)\delta_{n',\hat{n}(\Gamma(\tmid))}
\mathcal{T}_{n}(\hat{\Gamma}_<^\dagger) \nonumber\\
&=Z(n')\int \mathcal{D}\hat{\Gamma}_<^\dagger~ \frac{e^{-\beta H(\Gamma(\tmid);n')}}{ Z(n')}
\delta_{n',\hat{n}(\Gamma(\tmid))}
\mathcal{T}_{n}(\hat{\Gamma}_<^\dagger) \nonumber\\
&=Z(n')\int d\Gamma^{n,n'-n}
~\delta_{n',\hat{n}(\Gamma)}
\frac{e^{-\beta H(\Gamma;n')}}{Z(n')}.
\label{e:formula1-right0}
\end{align}
The last line of \eqref{e:formula1-right0} is not a path integral because \(Z(n')\) is constant over the time interval of the path \(\hat{\Gamma}_>^\dagger\),
and the integrand's observable, Kronecker's delta, depends only on \(\tmid\).
Using the definition of \(\rho(n|k)\) in \eqref{e:rho-def}, we further transform \eqref{e:formula1-right0} as
\begin{align}
&\mathrm{(RHS)}=Z(n')\rho(n'|n').
\label{eEM:LDB3}
\end{align}

The path integral on the left-hand side of \eqref{eEM:LDB2} is written as
\begin{align}
\int d\Gamma &\delta_{n',\hat{n}(\Gamma)} e^{-\beta W_{n,n'}} 
\int \mathcal{D}\hat{\Gamma}_<\int \mathcal{D}\hat{\Gamma}_>\nm
&
e^{-\beta H(\Gamma(\tini);n)}\mathcal{T}_{n}(\hat{\Gamma}_<)\delta(\Gamma-\Gamma(\tmid))
\mathcal{T}_{n'}(\hat{\Gamma}_>)
\end{align}
by applying \(\int d\Gamma \delta(\Gamma-\Gamma(\tmid))=1\).
This is further transformed as
\begin{align}
=\int d\Gamma &\delta_{n',\hat{n}(\Gamma)} e^{-\beta W_{n,n'}}e^{-\beta H(\Gamma;n)}\nm
~~~~~&\times \int \mathcal{D}\hat{\Gamma}_>\delta(\Gamma-\Gamma(\tmid))\mathcal{T}_{n'}(\hat{\Gamma}_>).
\end{align}
We then obtain
\begin{align}
&\mathrm{(LHS)}=\int d\Gamma \delta_{n',\hat{n}(\Gamma)} e^{-\beta W_{n,n'}}
e^{-\beta H(\Gamma;n)}.
\label{eEM:LDB3-L}
\end{align}

Equation~\eqref{eEM:LDB3-L} is expressed as the integral of \(\Gamma\), i.e., \(\Gamma^\subIN\) for the system of \(n\) C molecules, however, 
the natural expression of the microstates with \(n\) molecules of \ce{C} is \(\Gamma^{n,0}=(\Gamma_\subA^{=},\Gamma_\subA^{\neq},\Gamma_\subB^{=},\Gamma_\subB^{\neq})\) rather than \(\Gamma^\subIN\).
They are connected by \(\Gamma_\subA^{\neq}=(\Gamma_\subA^{\neq =},\Gamma_\subA^{\neq\neq})\),
\(\Gamma_\subB^{\neq}=(\Gamma_\subB^{\neq =},\Gamma_\subB^{\neq\neq})\), \(\Gamma_\subA^{=}=\Gamma_\subA^{==}\), and \(\Gamma_\subB^{=}=\Gamma_\subB^{==}\),
and therefore, changing the notations from \(\Gamma^\subIN\) to \(\Gamma^{n,0}\) in \eqref{eEM:LDB3-L},
we need to consider the difference in distinguishability in \(\Gamma_\subA^{\neq}\) and \(\Gamma_\subB^{\neq}\).
Using \(H(\Gamma^\subIN;n)=H(\Gamma^{n,0};n)\)
under the constraint \(\hat{n}(\Gamma^\subIN)=\hat{n}(\Gamma^{n,0})=n'\),
we have
\begin{align}
&\mathrm{(LHS)}=\left[\frac{(\nA-n)!}{(\nA-n')!(n'-n)! } \frac{(\nB-2n)!}{(\nB-2n')!~2^{n'-n}}\right]^{-1}\nm
&~~\int d\Gamma^{n,0}~
\delta_{n',\hat{n}(\Gamma^{n,0})}
e^{-\beta W_{n,n'}}
e^{-\beta H(\Gamma^{n,0};n)},
\label{eEM:integral}
\end{align}
whose integral becomes \(Z(n)\rho(n'|n)\bbkt{e^{-\beta W_{n,n'}}}_{n'}\) by using \eqref{e:rho-def}.
Thus, the equality \eqref{eEM:LDB2} is integrated as
\begin{align}
e^{-\beta F(n)}\rho(n'|n)\bbkt{e^{-\beta W_{n,n'}}}_{n'}
=
e^{-\beta F(n,n'-n)}\rho(n'|n').
\end{align}

\section{Effective Potential}
\label{sec:EM_effectiveF}

The effective potential for the trimer number in the presence of \(k\) stable \ce{C} molecules is defined as
\begin{align}
\mathcal{F}(n;k)
\equiv - \kB T \ln \frac{Z(n|k)}{(\nA-k)! k! (\nB-2k)! 2^k },
\end{align}
where
\begin{align}
Z(n|k)
= \int d\Gamma_\subA d\Gamma_\subB~
\delta_{n,\hat{n}(\Gamma_\subA,\Gamma_\subB)}
e^{-\beta H(\Gamma_\subA, \Gamma_\subB;k)}.
\end{align}
Since \(\rho(n|k)=Z(n|k)/Z(k)\), this effective potential is written as
\begin{align}
\mathcal{F}(n;k)=F(k)-\kB T\ln\rho(n|k).
\end{align}
For a staged conversion \(0\to k\to k'\), the corresponding history-dependent effective potential is
\begin{align}
\mathcal{F}(n;k,k'-k)=F(k,k'-k)-\kB T\ln\rho(n|k').
\end{align}
Thus \(\mathcal{F}(n;k,k'-k)\) has the same \(n\)-dependence as \(\mathcal{F}(n;k')\). Their difference is an \(n\)-independent offset. Taking the ordinary final ensemble as the reference gives
\begin{align}
\mathcal{F}(n;k')-\mathcal{F}(n;k,k'-k)
&= F(k')-F(k,k'-k) \nonumber\\
&= \kB T\ln\frac{k'!}{k!(k'-k)!}.
\end{align}
This positive offset is \(\Fgain\) in \eqref{e:Fgain_definition}: the history-resolved final ensemble is lower than the ordinary final ensemble by this amount. The offset comes from statistical counting, not from an additional force acting on the products.

\begin{figure}[tb]
\begin{center}
\includegraphics[width=6cm]{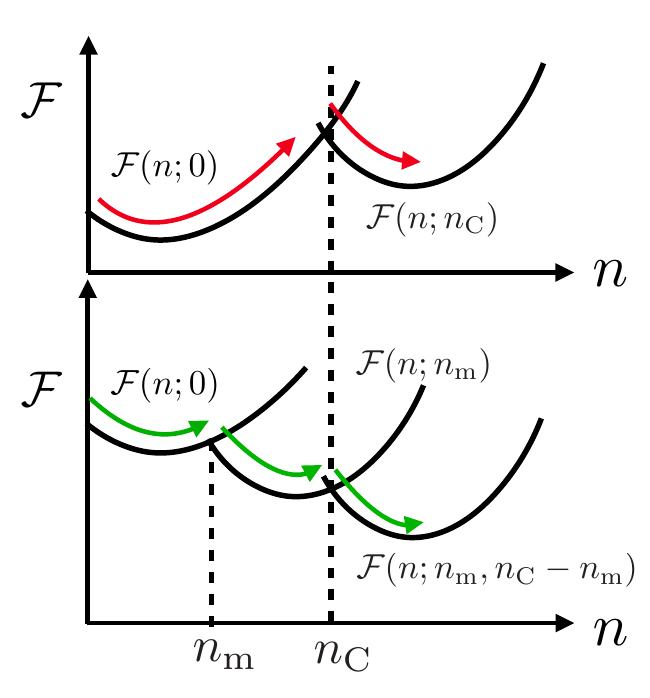}
\end{center}
\caption{
Effective-potential representation of the history-dependent Gibbs factorial.
The single and staged protocols assign different statistical counting to the final ensemble, represented by \(\mathcal{F}(n;\nC)\) and \(\mathcal{F}(n;\nint,\nC-\nint)\), respectively. The offset \(\mathcal{F}(n;\nC)-\mathcal{F}(n;\nint,\nC-\nint)=\Fgain\) is the free-energy representation of the Gibbs-factorial contribution.}
\label{fig:EM_effective_potential}
\end{figure}

Figure~\ref{fig:EM_effective_potential} illustrates this construction.
In the single protocol, the system first fluctuates on the initial landscape
\(\mathcal{F}(n;0)\) until the threshold \(n=\nC\) is reached, and the
resulting \(\nC\) complexes are then stabilized as \ce{C} molecules.
In the staged protocol, the system first reaches \(n=\nint\), stabilizes
these complexes, and then fluctuates from the landscape with \(\nint\)
stable products until the final threshold \(n=\nC\) is reached.
The two protocols therefore end with the same number of stable products but with different statistical counting. The vertical offset in Fig.~\ref{fig:EM_effective_potential} is the positive difference \(\mathcal{F}(n;\nC)-\mathcal{F}(n;\nint,\nC-\nint)\), i.e., the free-energy form of the Gibbs-factorial contribution.

\begin{thebibliography}{99}

\bibitem{Gibbs}
J. W. Gibbs,
{\it Elementary Principles in Statistical Mechanics} 
(Yale University Press, New Haven, 1902).

\bibitem{Pathria}
R. K. Pathria and P. D. Beale,
{\it Statistical Mechanics}, 3rd ed.
(Elsevier, Amsterdam, 2011).

\bibitem{Murashita-Ueda}
Y. Murashita and M. Ueda,
Gibbs paradox revisited from the fluctuation theorem with absolute irreversibility,
Phys. Rev. Lett. 118, 060601 (2017). 
DOI: 10.1103/PhysRevLett.118.060601

\bibitem{Yoshida-Nakagawa}
A. Yoshida and N. Nakagawa,
Work relation for determining the mixing free energy of small-scale mixtures,
Phys. Rev. Res. 4, 023119 (2022). 
DOI: 10.1103/PhysRevResearch.4.023119

\bibitem{SHNY}
S.-i. Sasa, K. Hiura, N. Nakagawa, and A. Yoshida,
Quasi-static decomposition and the Gibbs factorial in small thermodynamic systems,
J. Stat. Phys. 189, 31 (2022). 
DOI: 10.1007/s10955-022-02991-7

\bibitem{Perutz}
M. F. Perutz, 
Mechanisms of cooperativity and allosteric regulation in proteins, 
Q. Rev. Biophys. 22, 139-237 (1989). 
DOI: 10.1017/s0033583500003826

\bibitem{Monod}
J. Monod, J. Wyman, and J. P. Changeux, 
On the nature of allosteric transitions: a plausible model, 
J. Mol. Biol. 12, 88-118 (1965). 
DOI: 10.1016/s0022-2836(65)80285-6

\bibitem{Koshland}
D. E. Koshland Jr., G. N\'{e}methy, and D. L. Filmer, 
Comparison of experimental binding data and theoretical models in proteins containing subunits, 
Biochemistry 5, 365-385 (1966). 
DOI: 10.1021/bi00865a047

\bibitem{Cui}
Q. Cui and M. Karplus, 
Allostery and cooperativity revisited, 
Protein Sci. 17, 1295-1307 (2008). 
DOI: 10.1110/ps.03259908

\bibitem{Martinez-Frenkel}
F. J. Martinez-Veracoechea and D. Frenkel,
Designing super selectivity in multivalent nano-particle binding,
Proc. Natl. Acad. Sci. U.S.A. 108, 10963-10968 (2011).
DOI: 10.1073/pnas.1105351108

\bibitem{Liu2020}
M. Liu, A. Apriceno, M. Sipin, E. Scarpa, L. Rodriguez-Arco, A. Poma, G. Marchello, G. Battaglia, and S. Angioletti-Uberti,
Combinatorial entropy behaviour leads to range selective binding in ligand-receptor interactions,
Nat. Commun. 11, 4836 (2020).
DOI: 10.1038/s41467-020-18603-5

\bibitem{SM}
See Supplemental Material at [URL will be inserted by publisher] for the model potential, simulation parameters, criteria for identifying transient complexes, time-scale analysis, waiting-time measurements, and supplemental numerical data, which includes Refs. \cite{WCA, SW, Nakamura}.

\bibitem{WCA}
J. D. Weeks, D. Chandler, and H. C. Andersen, 
Role of repulsive forces in determining the equilibrium structure of simple liquids, 
J. Chem. Phys. 54, 5237 (1971). 
DOI: 10.1063/1.1674820

\bibitem{SW}
F. H. Stillinger and T. A. Weber, 
Computer simulation of local order in condensed phases of silicon, 
Phys. Rev. B 31, 5262 (1985). 
DOI: 10.1103/PhysRevB.31.5262

\bibitem{Nakamura}
T. Nakamura, in preparation.

\bibitem{Jarzynski}
C. Jarzynski, 
Nonequilibrium equality for free energy differences,
Phys. Rev. Lett. 78, 2690-2693 (1997).
DOI: 10.1103/PhysRevLett.78.2690

\bibitem{Crooks}
G. E. Crooks,
Entropy production fluctuation theorem and the nonequilibrium work relation for free energy differences,
Phys. Rev. E 60, 2721-2726 (1999).
DOI: 10.1103/PhysRevE.60.2721

\bibitem{Seifert}
U. Seifert, 
Stochastic thermodynamics, fluctuation theorems and molecular machines, 
Rep. Prog. Phys. 75, 126001 (2012). 
DOI: 10.1088/0034-4885/75/12/126001

\bibitem{SagawaUeda} 
T. Sagawa and M. Ueda, 
Generalized Jarzynski equality under nonequilibrium feedback control, 
Phys. Rev. Lett. 104, 090602 (2010).
DOI: 10.1103/PhysRevLett.104.090602

\bibitem{Dunn2015Nature}
K. E. Dunn, F. Dannenberg, T. E. Ouldridge, M. Kwiatkowska, A. J. Turberfield, and J. Bath,
Guiding the folding pathway of DNA origami,
Nature 525, 82-86 (2015).
DOI: 10.1038/nature14860

\bibitem{Wang2012}
Y. Wang, Y. Wang, D. R. Breed, V. N. Manoharan, L. Feng, A. D. Hollingsworth, M. Weck, and D. J. Pine,
Colloids with valence and specific directional bonding,
Nature 491, 51-55 (2012).
DOI: 10.1038/nature11564

\bibitem{DiMichele2013}
L. Di Michele, F. Varrato, J. Kotar, S. H. Nathan, G. Foffi, and E. Eiser,
Multistep kinetic self-assembly of DNA-coated colloids,
Nat. Commun. 4, 2007 (2013).
DOI: 10.1038/ncomms3007

\bibitem{Gong2017}
Z. Gong, T. Hueckel, G.-R. Yi, and S. Sacanna,
Patchy particles made by colloidal fusion,
Nature 550, 234-238 (2017).
DOI: 10.1038/nature23901

\bibitem{Swan2014}
J. W. Swan, J. L. Bauer, Y. Liu, and E. M. Furst,
Directed colloidal self-assembly in toggled magnetic fields,
Soft Matter 10, 1102-1109 (2014).
DOI: 10.1039/C3SM52663A
\end{thebibliography}
\end{document}


\title{Supplemental Material for ``Gibbs Factorials Become Kinetic in History-Dependent Reactions''}

\author{Nanako Hirano}
\affiliation{Department of Physics, Ibaraki University, Mito 310-8512, Japan}

\author{Akira Yoshida}
\affiliation{Department of Physics, Ibaraki University, Mito 310-8512, Japan}
\affiliation{Department of Physics, Kyoto University, Kyoto 606-8502, Japan}

\author{Takenobu Nakamura}
\affiliation{National Institute of Advanced Industrial Science and Technology (AIST), Tsukuba 305-8568, Japan}

\author{Naoko Nakagawa}
\affiliation{Department of Physics, Ibaraki University, Mito 310-8512, Japan}

\maketitle

This Supplemental Material provides numerical and technical details supporting the molecular-dynamics verification in the main text.
The fluctuation-theorem derivation and the effective-potential interpretation are given in the End Matter of the main text and are not repeated here.
Instead, the focus is on the model Hamiltonian, the equilibrium sampling, the identification of transient \ce{AB2} complexes, the time-scale hierarchy, and the numerical evaluation of the probabilities and waiting times entering the kinetic advantage factor \(\Ac\).

\section{Model Hamiltonian and interaction potentials}
\label{s:Hamiltonian}

\begin{figure}[tb]
\centering
\includegraphics[width=0.95\textwidth]{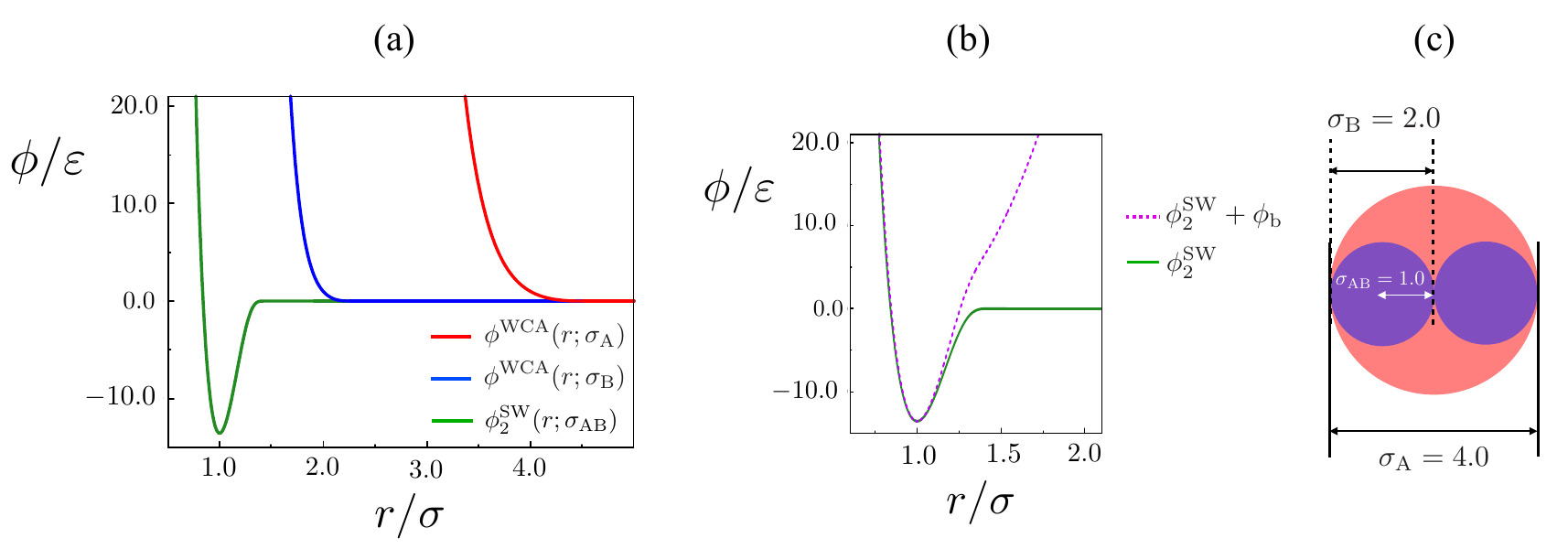}
\caption{
(a) Interaction potentials \(\phi^{\mathrm{WCA}}(r;\sigma_{\subA})\), \(\phi^{\mathrm{WCA}}(r;\sigma_{\subB})\), and \(\phi^{\mathrm{SW}}_2(r;\sigma_{\subAB})\), with \(\sigma_{\subA}=4.0\sigma\), \(\sigma_{\subB}=2.0\sigma\), \(\sigma_{\subAB}=1.0\sigma\), and \(\theta_0=\pi\).
\(\phi^{\mathrm{SW}}_3=0\) when \(\theta=\theta_0\) or \(r>1.5\sigma\), and \(\phi^{\mathrm{SW}}_3>0\) otherwise.
(b) Comparison of \(\phi^{\mathrm{SW}}_2(r;\sigma_{\subAB})\) with \(\phi^{\mathrm{SW}}_2(r;\sigma_{\subAB})+\phi_{\mathrm b}(r)\).
\(\phi_{\mathrm b}\) does not significantly modify the attractive interaction near \(r/\sigma\simeq1\), but becomes steep with increasing \(r\), ensuring that \ce{A} and \ce{B} remain bound after conversion.
In contrast, \(\phi^{\mathrm{SW}}_2(r;\sigma_{\subAB})\) becomes flat for \(r/\sigma_{\subAB}>1.5\), so the transient \ce{A}--\ce{B} association is not maintained under large fluctuations.
(c) Schematic structure of the transient \ce{AB2} complex.
}
\label{fig:potential_MD}
\end{figure}

For numerical demonstrations, we use a Hamiltonian for a three-dimensional mixture of \(\nA\) molecules of species \ce{A} and \(\nB\) molecules of species \ce{B},
\begin{align}
H_0(\Gamma_{\subA},\Gamma_{\subB})
&= \sum_{i=1}^{\nA} \frac{ |\bm{p}^{\subA}_i |^2 }{2m_{\subA}}
+ \sum_{j=1}^{\nB} \frac{ |\bm{p}^{\subB}_j |^2 }{2m_{\subB}}
+ \sum_{i_1=1}^{\nA} \sum_{i_2>i_1}^{\nA} \phi^{\mathrm{WCA}}(|\bm{r}_{i_1}^{\subA}-\bm{r}_{i_2}^{\subA}|;\sigma_{\subA})
+ \sum_{j_1=1}^{\nB} \sum_{j_2>j_1}^{\nB} \phi^{\mathrm{WCA}}(|\bm{r}_{j_1}^{\subB}-\bm{r}_{j_2}^{\subB}|;\sigma_{\subB}) \nonumber\\
&+ \sum_{i=1}^{\nA} \sum_{j=1}^{\nB} \phi^{\mathrm{SW}}_2(|\bm{r}_{i}^{\subA}-\bm{r}_{j}^{\subB}|;\sigma_{\subAB})
+ \sum_{i=1}^{\nA} \sum_{j_1=1}^{\nB} \sum_{j_2>j_1}^{\nB} \phi^{\mathrm{SW}}_3(|\bm{r}_{i}^{\subA}-\bm{r}_{j_1}^{\subB}|, |\bm{r}_{i}^{\subA}-\bm{r}_{j_2}^{\subB}|, \theta_{j_1ij_2};\sigma_{\subAB},\theta_0).
\label{e:H0-supp}
\end{align}
Here \(m_{\subA}=m_{\subB}=m\).
The Weeks-Chandler-Andersen interaction \cite{SM:WCA} is
\begin{align}
\phi^{\mathrm{WCA}}(r;\sigma)
=4\eunit\left[\left(\frac{\sigma}{r}\right)^{12}
-\left(\frac{\sigma}{r}\right)^6\right]\Theta(2^{1/6}\sigma-r),
\label{e:WCA}
\end{align}
describing the two-body softcore-repulsive interaction with diameter \(\sigma\) and step function \(\Theta(r)\).
The two- and three-body Stillinger-Weber interactions \cite{SM:SW} are
\begin{align}
\phi^{\mathrm{SW}}_2(r;\sigma)
&=A\eunit\left[B\left(\frac{\sigma}{r}\right)^p
-\left(\frac{\sigma}{r}\right)^q\right]
\exp\left(\frac{\sigma}{r-1.5\sigma}\right)\Theta(1.5\sigma-r),\\
\phi^{\mathrm{SW}}_3(r,r',\theta;\sigma,\theta_0)
&=\lambda\eunit(\cos\theta-\cos\theta_0)^2
\exp\left(\frac{\sigma}{r-1.5\sigma}\right)
\exp\left(\frac{\sigma}{r'-1.5\sigma}\right)
\Theta(1.5\sigma-r)\Theta(1.5\sigma-r').
\label{e:SW}
\end{align}
The dimensionless parameters are \(A=200\), \(B=0.50\), \(p=4.0\), \(q=0.0\), and \(\lambda=100\).
In the last term of Eq.~\eqref{e:H0-supp}, \(\theta_{j_1ij_2}\) is the angle formed by the two vectors
\(\bm r_i^{\subA}-\bm r_{j_1}^{\subB}\) and \(\bm r_i^{\subA}-\bm r_{j_2}^{\subB}\) at the \(\ce{A}\) particle.
\(\phi^{\mathrm{SW}}_2(r)\) is repulsive for \(r\le\sigma_{\subAB}\) and attractive for \(\sigma_{\subAB}<r\le1.5\sigma_{\subAB}\).
\(\phi^{\mathrm{SW}}_3\) stabilizes the angle \(\theta_{j_1ij_2}\) at \(\theta_0\) and acts only when two \ce{B} molecules are near the same \ce{A} molecule.
We set \(\sigma_{\subA}=4.0\lunit\), \(\sigma_{\subB}=2.0\lunit\), \(\sigma_{\subAB}=1.0\lunit\), and \(\theta_0=\pi\).
This model follows the molecular-design study in Ref.~\cite{SM:Nakamura}.
The attractive part of \(\phi_2^{\mathrm{SW}}\) transiently traps two \ce{B} molecules around one \ce{A} molecule, producing the reversible association \(\ce{A + 2B <=> AB2}\).
The interaction potentials between two \ce{A}, between two \ce{B}, and between \ce{A} and \ce{B} are plotted in Fig.~\ref{fig:potential_MD}(a) for \(\theta=\theta_0\).

The irreversible operation \(\ce{AB2 -> C}\) is represented by adding a harmonic binding potential between the \ce{A} particle and the two \ce{B} particles in a transient complex,
\begin{align}
\phi_{\mathrm{b}}(\bm{r}_{2i-1}^{\subB},\bm{r}_{2i}^{\subB};\bm{r}_i^{\subA})
=\frac{\ep_\mathrm{b}}{2}\sum_{j=2i-1}^{2i}
\left(|\bm{r}^{\subA}_{i}-\bm{r}^{\subB}_{j}|-\sigma\right)^2,
\label{e:bond-potential-supp}
\end{align}
with \(\ep_\mathrm{b}=40\eunit/\sigma^2\).
The bond length is chosen at the minimum of the attractive \ce{A}--\ce{B} interaction, \(l=\sigma\).
The value of \(\ep_\mathrm{b}\) is chosen so that \(\phi_{\mathrm b}\) minimally affects the properties around the minimum of \(\phi_2^{\mathrm{SW}}(r)\), while preventing dissociation after conversion.
The added potential stabilizes the product \ce{C} but does not remove the underlying vibrational degrees of freedom; all original nonbonded interactions remain present.
Thus the harmonic binding potential selects the \ce{A}--\ce{B} bond distances relevant to the conversion event, but it does not constrain the remaining molecular motions or reduce the phase-space dimension.
In this sense \ce{C} is a stabilized version of the same trimer structure, not a new mechanically cooperative product.

Hereafter we use \(\sigma=1\), \(\eunit=1\), \(m=1\), and \(\kB=1\).

\section{Molecular-dynamics setup and equilibrium sampling}
\label{s:MD_setup}

We perform molecular-dynamics simulations in an \(NVT\) ensemble using a Nose-Hoover thermostat.
The system is placed in a cubic periodic box of side length \(L_x=L_y=L_z=45\), and we set \(\nA:\nB=1:2\) with \(\nA=108\) and \(\nB=216\).
The equations of motion are integrated using the velocity-Verlet method with time step \(dt=0.005\).
The initial configuration is a regular lattice of \ce{AB2} complexes, \(\hat n(t=0)=\nA\), and initial velocities are sampled from the Maxwell distribution.

For equilibrium measurements of probability distributions, we use a conservative equilibration time much longer than the microscopic relaxation and transient-complex lifetime discussed in \hyperref[s:timescale]{Sec.~VII}.
The system is equilibrated up to \(t=1.0\times10^7\).
Sampling is then performed from \(t=1.0\times10^7\) to \(2.0\times10^7\) at intervals of 50 time units, giving \(2.0\times10^5\) samples for each parameter set.
Measurements of lifetime and waiting time require finer temporal information and are described separately in Secs.~\hyperref[s:criterion]{IV}--\hyperref[s:wtime]{VIII}.

The main text uses Langevin dynamics for the path-probability derivation because local detailed balance takes the standard form there.
This does not constrain the equilibrium distributions used in the numerical test: the probabilities \(\rho(n|k)\) are canonical equilibrium quantities and are independent of the thermostat once equilibration is achieved.

\section{Thermodynamic consistency of transient \ce{AB2} and stable \ce{C}}
\label{s:eos}

The construction in the main text assumes that the stabilized product \ce{C} and a transient \ce{AB2} complex are indistinguishable for the coarse thermodynamic observables relevant here, apart from stability.
We check this by comparing virial pressures.
This check is useful because association and dissociation of \ce{AB2} complexes change the number of free particles contributing to pressure.
At low temperature, where complex formation dominates, nearly all \ce{A} and \ce{B} molecules form \ce{AB2} complexes, and the pressure should match that of a reference system in which all trimers are stabilized as \ce{C}.

\begin{figure}[tb]
\centering
\includegraphics[width=7cm]{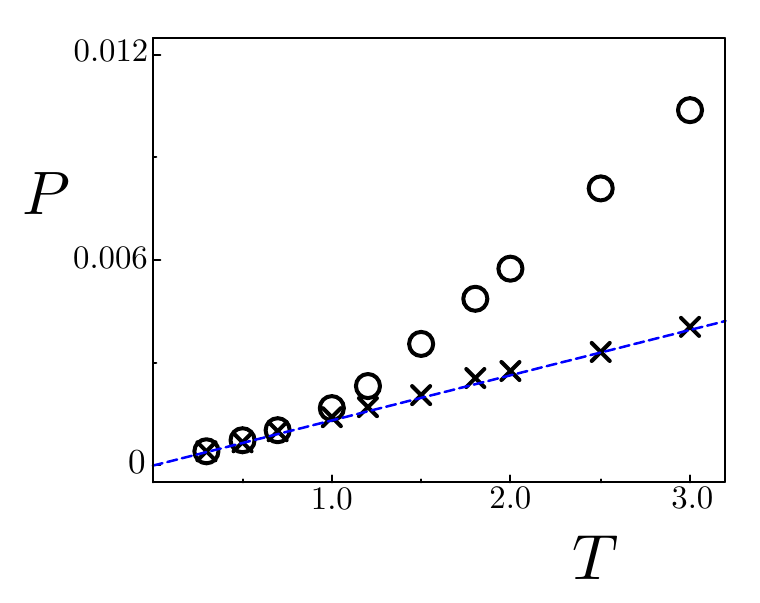}
\caption{
Temperature dependence of the virial pressure \(P\) for the reversible \ce{A}--\ce{B} system with \(k=0\) and the fully stabilized reference system with \(k=\nA\).
The dashed curve is the second-virial expression in Eq.~\eqref{e:eos_C}.
}
\label{fig:press_temp}
\end{figure}

The pressure is plotted in Fig.~\ref{fig:press_temp}.
The reversible system shows a crossover around \(T=1.0\): in both the low-temperature range \(0.3\le T<1.0\) and the high-temperature range \(1.0<T\le3.0\), the pressure is approximately linear in \(T\), but with different slopes.
In the low-temperature regime, the pressure of the reversible system is close to that of the fully stabilized \ce{C} reference system.
This agreement supports the interpretation that the low-temperature reversible system is a dilute gas of trimers.
It also checks that the chosen \(\ep_{\mathrm b}\) and \(l\) do not significantly alter macroscopic observables such as pressure.
The dashed curve is
\begin{align}
P=\frac{\nA \kB T}{V}\left(1+b_2\frac{\nA}{V}\right),
\label{e:eos_C}
\end{align}
with \(b_2=97\).
The crossover around \(T\simeq1.0\) indicates the onset of dissociation into free \ce{A} and \ce{B} molecules.

\section{Identifying transient \ce{AB2} complexes}
\label{s:criterion}

The observable \(\hat n\) counts stable \ce{C} molecules and transient \ce{AB2} complexes.
For a given \ce{A} particle, the geometric count of nearby \ce{B} particles is
\begin{align}
\hat q_i(\Gamma)
=\sum_{j=1}^{\nB}\Theta(r_0-|\bm r_i^\subA-\bm r_j^\subB|)
\prod_{i'\neq i}\left[1-\Theta(r_0-|\bm r_{i'}^\subA-\bm r_j^\subB|)\right].
\label{e:qi_supp}
\end{align}
The \(i\)-th \ce{A} particle satisfies the geometric condition for being in \ce{AB2} when \(\hat q_i=2\).
The total trimer number in the presence of \(k\) pre-existing \ce{C} molecules is then
\begin{align}
\hat n(\Gamma;k)=k+\sum_{i=k+1}^{\nA}\delta_{\hat q_i(\Gamma),2}.
\label{e:n_supp_raw}
\end{align}

The geometric criterion alone can count brief close approaches in which two \ce{B} molecules pass within \(r_0\) of an \ce{A} particle without forming a persistent complex.
The probability distributions \(\rho(n|k)\) and the fluctuation-theorem argument use the instantaneous phase-space observable \(\hat n(\Gamma;k)\) defined above.
For waiting times, however, a successful arrival at the target is counted only when the transient complex persists beyond a microscopic recrossing event.
This persistence requirement is an operational definition of the reaction event in the waiting-time measurement; it does not redefine the equilibrium probabilities \(\rho(n|k)\).
In the final waiting-time analysis, this persistence requirement is implemented by the time-averaged-distance trimer number \(\hat n_\Delta(t;k)\) defined in \hyperref[s:timescale]{Sec.~VII}.
The corresponding waiting-time intervals are specified operationally in \hyperref[s:wtime]{Sec.~VIII}.

The choice \(r_0=1.5\) is determined from the dependence of the mean complex number on \(r_0\).
First, \(r_0\) should be greater than \(\sigma_{\subAB}\) and less than \((\sigma_{\subA}+\sigma_{\subB})/2\), i.e., \(1.0<r_0<3.0\).
If \(r_0<\sigma_{\subAB}\), almost no \ce{AB2} complexes are counted, because the cutoff lies inside the repulsive core.
The average \(n\equiv\bar n\) is a function of \(r_0\) and \(T\).
At \(T=0.5\), which is low enough for almost all molecules to associate as \ce{AB2}, the expected value \(n\simeq\nA\) is obtained for \(r_0\gtrsim1.2\).
At \(T=2.5\), where \ce{AB2} complexes coexist with free \ce{A} and \ce{B}, \(n\) changes stepwise near \(r_0=1.2\), remains almost unchanged up to \(r_0\simeq1.7\), and then starts to increase.
The increase at larger \(r_0\) is attributed to misidentification: two free \ce{B} molecules that are not trapped by the attractive part of \(\phi^{\mathrm{SW}}_2\) can be counted as part of a complex when they happen to be near the same \ce{A}.
We therefore adopt \(r_0=1.5\).

\begin{figure}[tb]
\centering
\includegraphics[width=14cm]{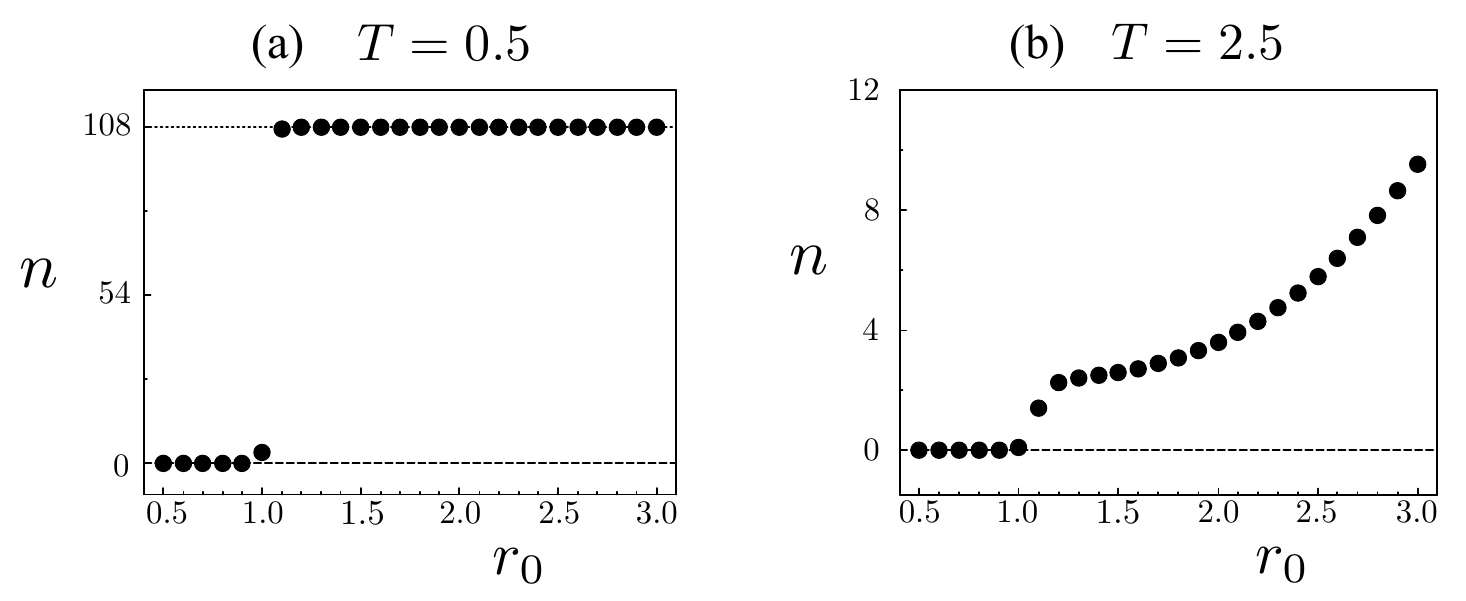}
\caption{
Mean number of \ce{A} particles identified as belonging to \ce{AB2} complexes as a function of \(r_0\) for \(k=0\).
(a) \(T=0.5\), where almost all trimers are formed.
(b) \(T=2.5\), used in the numerical tests of the main text.
The plateau around \(r_0=1.5\) motivates the criterion used in the simulations, while the increase at larger \(r_0\) reflects misidentification of free \ce{B} molecules.
}
\label{fig:n_r0}
\end{figure}

\begin{figure}[tb]
\centering
\includegraphics[width=7cm]{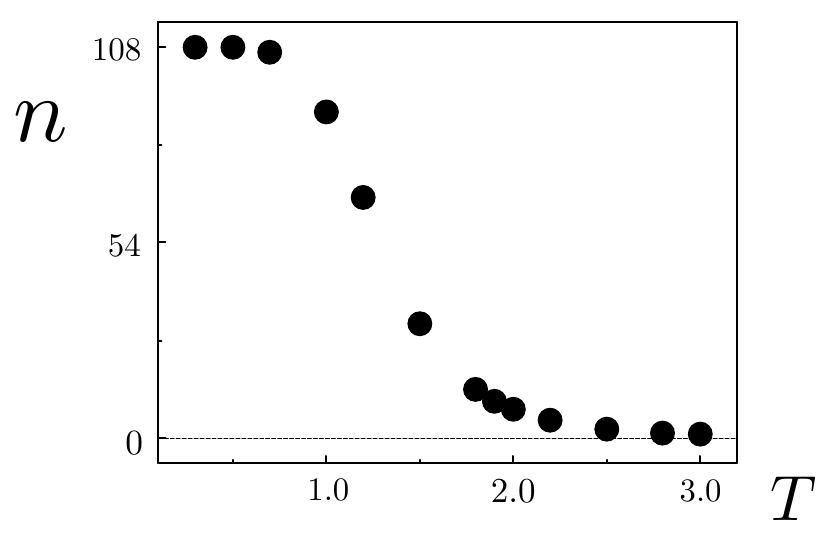}
\caption{
Mean trimer number \(n\) as a function of temperature at \(r_0=1.5\) and \(k=0\).
The system is almost fully associated for \(T<1.0\), while transient \ce{AB2} complexes coexist with free \ce{A} and \ce{B} molecules in the range \(1.0\lesssim T<3.0\).
}
\label{fig:n_lowtemp}
\end{figure}

\section{Choice of temperature and accessible target sizes}
\label{s:choice_T}

The numerical tests in the main text are performed at \(T=2.5\).
As shown in Fig.~\ref{fig:n_lowtemp}, this temperature lies in the regime where only a small number of transient \ce{AB2} complexes remain in equilibrium.
It is therefore high enough that the threshold events are nontrivial rare events, but low enough that the relevant tail probabilities can still be sampled.
We also restrict the final target to \(\nC\le10\) in the numerical test.
This restriction is imposed only by numerical accessibility and is not part of the theoretical result.
To evaluate
\begin{align}
\Ac(\nint,\nC)
=\frac{\rho(\nC|\nint)}{\rho(\nC|0)}
\frac{\rho(\nint|0)}{\rho(\nint|\nint)},
\label{e:Ac_supp}
\end{align}
we need reliable estimates of four probabilities.
Let \(n_*(0)\) and \(n_*(\nint)\) be the most probable values of \(\rho(n|0)\) and \(\rho(n|\nint)\).
Since the tested regime satisfies \(\nC>\nint>n_*(0)\), and since the shift of the distribution gives \(\nC-n_*(\nint)<\nC-n_*(0)\), the probability \(\rho(\nC|0)\) usually lies deeper in the tail than \(\rho(\nC|\nint)\).
Thus the accuracy of \(\rho(\nC|0)\), together with that of \(\rho(\nint|\nint)\), is the limiting factor for evaluating \(\Ac\).

\begin{figure}[tb]
\centering
\includegraphics[width=13cm]{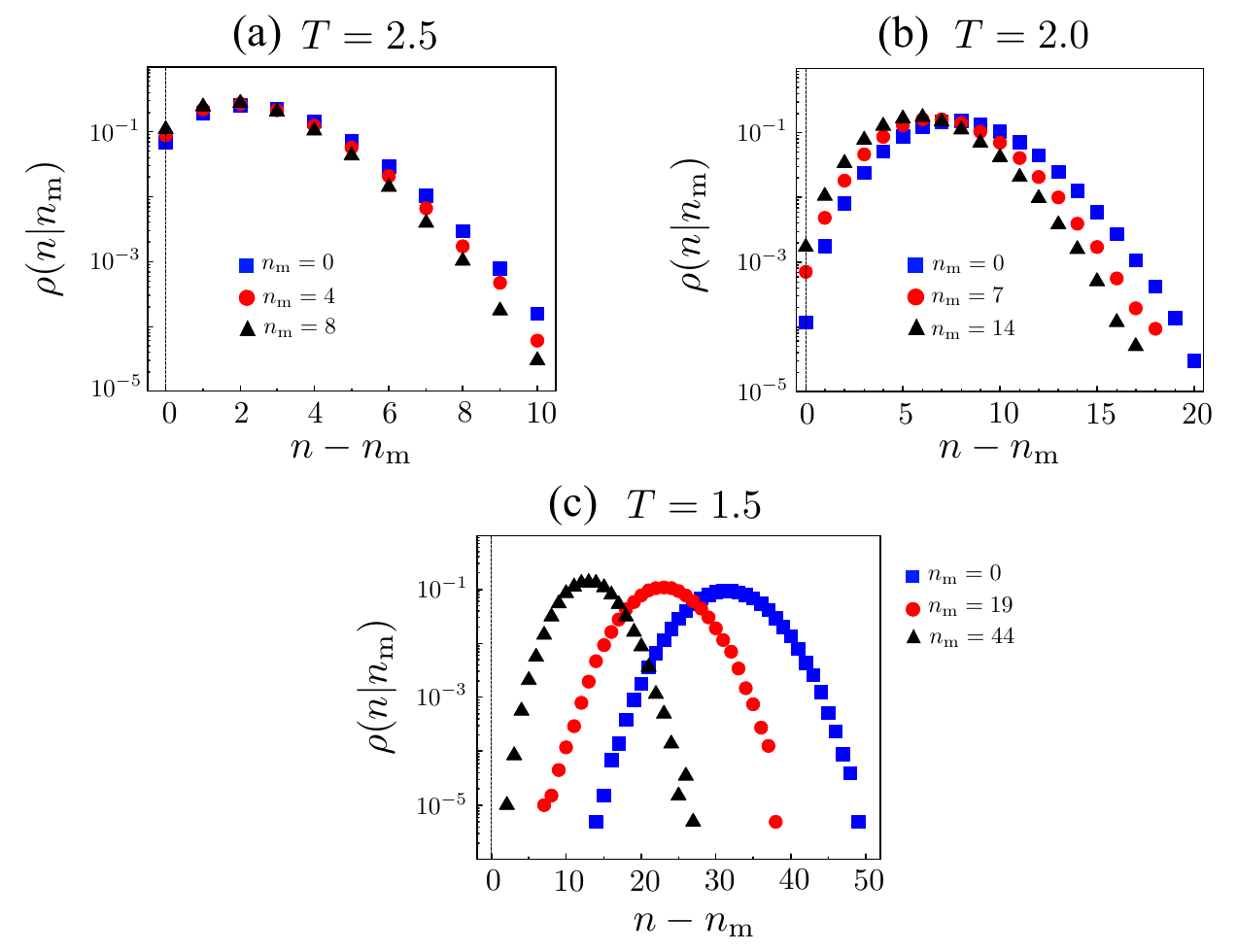}
\caption{
Probability distributions \(\rho(n|\nint)\) at three temperatures:
(a) \(T=2.5\) with \(\nint=0,4,8\),
(b) \(T=2.0\) with \(\nint=0,7,14\), and
(c) \(T=1.5\) with \(\nint=0,19,44\).
The comparison shows why \(T=2.5\) gives the best compromise between a rare-event regime and numerical accessibility for \(\nC\le 10\).
}
\label{fig:rho_T_nm}
\end{figure}

Figure~\ref{fig:rho_T_nm} compares the distributions at \(T=2.5\), \(2.0\), and \(1.5\).
The magnitude of \(\rho(\nint|\nint)\) can be read from the value at \(n-\nint=0\).
At \(T=1.5\), \(\rho(\nint|\nint)\) becomes too small for efficient sampling in the tested range.
At \(T=2.0\), evaluating larger \(\nint\) is already computationally demanding.
At \(T=2.5\), \(\rho(\nint|\nint)\) remains accessible and targets up to \(\nC\le10\) can be sampled with the available data.
The value \(\nC=10\) is large enough to demonstrate the factorial enhancement but small enough to keep \(\rho(\nC|0)\) statistically resolvable.

\section{Additional numerical check of \(\Ac\)}
\label{s:Ac_eval}

\begin{figure}[tb]
\centering
\includegraphics[width=16cm]{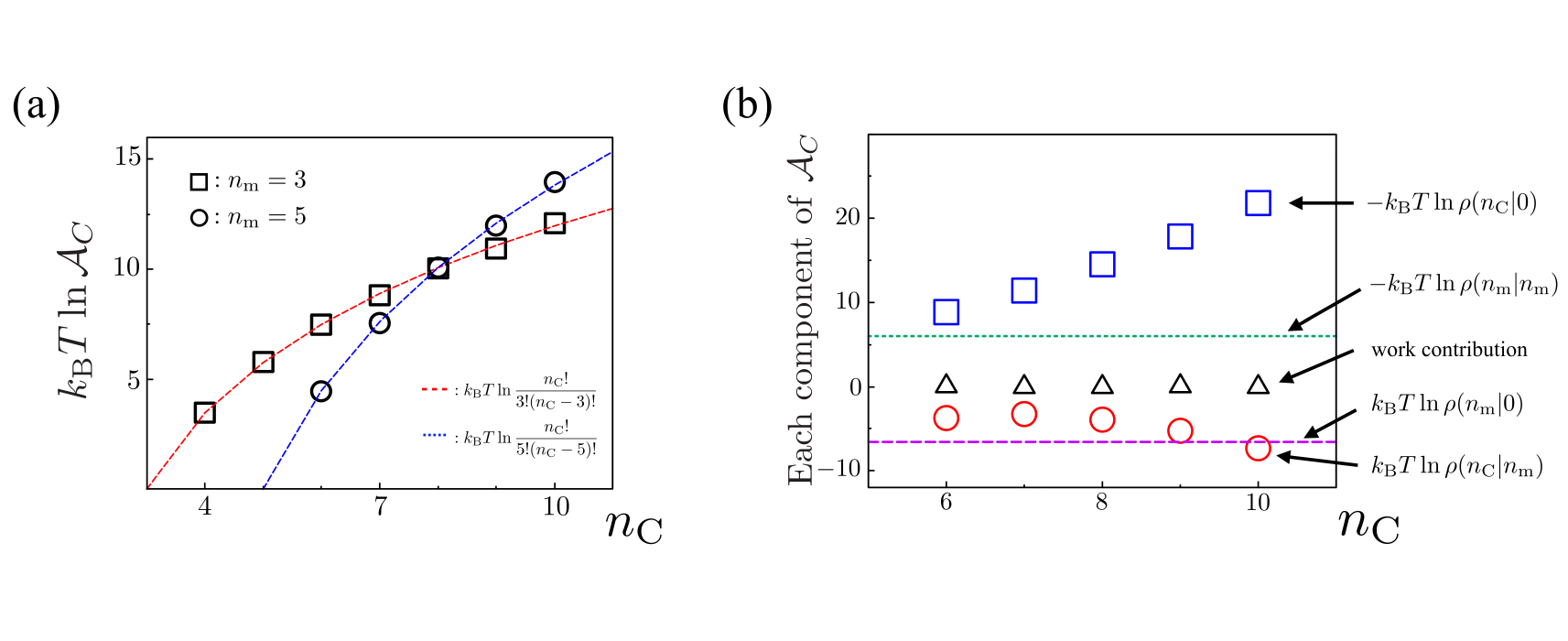}
\caption{
Supplemental numerical check of \(\Ac\).
(a) \(\kB T\ln\Ac\) as a function of \(\nC\) at fixed \(\nint=3\) and \(5\).
The guide lines are \(\kB T\ln[\nC!/\{\nint!(\nC-\nint)!\}]\).
(b) Probability components and the work contribution for the \(\nint=5\) data.
The work contribution is nearly vanishing and negligible on this scale.
}
\label{fig:deltaF}
\end{figure}

Figure~3 of the main text tests the Gibbs-factorial contribution by fixing \(\nC=10\) and varying the intermediate value \(\nint\).
As a supplemental check, Fig.~\ref{fig:deltaF} instead fixes \(\nint=3\) or \(5\) and varies \(\nC\).
The data provide a consistency check that the same factorial form describes the \(\nC\)-dependence,
\begin{align}
\kB T\ln\Ac(\nint,\nC)\simeq
\kB T\ln \frac{\nC!}{\nint!(\nC-\nint)!}.
\label{e:Ac_factorial_supp}
\end{align}
The decomposition in Fig.~\ref{fig:deltaF}(b) also shows that the work contribution is nearly vanishing and negligible on this scale.
Thus this figure is supplemental to the main numerical verification, rather than an independent derivation of \(\Ac\).

\section{Relaxation and time-scale hierarchy}
\label{s:timescale}

The staged-protocol analysis treats \(\hat n\), the total number of trimers, as the relevant slow order parameter for the present waiting-time measurement.
We define \(t_r\) as the conditional relaxation time within a sector of fixed \(\hat n\): after the system has a given number of trimers, the remaining microscopic degrees of freedom relax to the canonical distribution constrained by that value of \(\hat n\).
The relevant time-scale condition used in the main text is
\begin{align}
\tn>t_r,
\end{align}
where \(\tn\) is the correlation time of the coarse-grained trimer-number observable used in the waiting-time measurement.
Under this condition, microscopic relaxation is faster than the coarse-grained evolution of \(\hat n\), so the statistics of each stage can be described by the equilibrium probability \(\rho(n|k)\) and by a quasi-equilibrium coarse-grained dynamics of \(\hat n\).
This separation is conceptually distinct from the microscopic contact time introduced below, which is used to choose an averaging window that filters out brief close approaches in the waiting-time measurement.

We first estimate the local microscopic transport time from the mean free path \( \lambda \) of the dilute mixture.
For the numerical condition used in the main text, \(T=2.5\), \(L=45\), \(\nA=108\), and \(\nB=216\), the total number density is
\begin{align}
\rho=\frac{\nA+\nB}{L^3}=3.56\times10^{-3}.
\end{align}
The dilute hard-sphere estimate gives
\begin{align}
\lambda\simeq \frac{1}{\sqrt{2}\pi d_{\rm eff}^2\rho},
\qquad
\bar v=\sqrt{\frac{8T}{\pi}}=2.52,
\end{align}
where \(d_{\rm eff}\) is an effective collision diameter.
For \(d_{\rm eff}=1.5\), \(2.0\), and \(3.0\), the corresponding mean collision times
\begin{align}
t_{\rm mfp}\equiv \lambda/\bar v
\end{align}
are \(11.2\), \(6.3\), and \(2.8\), respectively.
Equivalently, the self-diffusion estimate \(D\simeq \bar v\lambda/3\) gives \(D=23.7\), \(13.3\), and \(5.9\).
The mean-free-path time \(t_{\rm mfp}\) is a lower-bound local transport scale, not the full conditional relaxation time \(t_r\).
For a conservative system-scale estimate, the ballistic crossing time is \(L/\bar v\simeq18\), while \(L^2/(6D)\) gives \(14\), \(25\), and \(57\) for the same three values of \(d_{\rm eff}\).
We therefore use
\begin{align}
10\lesssim t_r\lesssim60
\end{align}
as a conservative range for the conditional equilibration time at fixed \(\hat n\).

The time scale \(\tn\) is measured from the time-averaged-distance trimer number used in the waiting-time analysis.
For a time window \(I_\Delta(t)\) of width \(\Delta\), we define
\begin{align}
\bar r_{ij}^{\Delta}(t)
=
\frac{1}{\Delta}
\int_{I_\Delta(t)}
|\bm{r}_i^\subA(s)-\bm{r}_j^\subB(s)|\,ds ,
\end{align}
and
\begin{align}
\hat q_i^\Delta(t)
=
\sum^{\nB}_{j=1} \Theta \left( r_0 - \bar r_{ij}^{\Delta}(t) \right)
\prod_{i'\neq i}\left[1-\Theta \left( r_0 - \bar r_{i'j}^{\Delta}(t) \right) \right],
\qquad
\hat n_\Delta(t;k)
=
k+\sum_{i=k+1}^{\nA} \delta_{\hat q_i^\Delta(t), 2}.
\end{align}
For \(k=0\), this reduces to the total time-averaged count of transient \ce{AB2} complexes.
We note that the equilibrium probabilities \(\rho(n|k)\) use the instantaneous
structural criterion, whereas \(\hat n_\Delta\) is used only to define
successful events in the time-resolved measurement.
After equilibration up to \(t=1.0\times10^7\), we sampled
\(\hat n_\Delta(t;0)\) every \(5.0\) MD time units for a total duration of
\(2.0\times10^5\) at \(T=2.5\), \(k=0\), and \(\Delta=5.0\).
We compute the normalized autocorrelation function
\begin{align}
C_{\hat n}(s)
=
\frac{
\left\langle
\delta\hat n_\Delta(t+s;0)\delta\hat n_\Delta(t;0)
\right\rangle
}{
\left\langle
\left[\delta\hat n_\Delta(t;0)\right]^2
\right\rangle
},
\qquad
\delta\hat n_\Delta(t;0)=\hat n_\Delta(t;0)-\langle\hat n_\Delta(\cdot;0)\rangle .
\end{align}
The data give a \(1/e\) decay time \(1.9\times10^2\), an integrated correlation time \(2.1\times10^2\), and an exponential-fit time \(2.2\times10^2\).
We therefore use
\begin{align}
\tn\simeq 2\times10^2
\end{align}
as the characteristic time scale of the coarse-grained trimer-number process.
Together with \(10\lesssim t_r\lesssim60\), this supports \(\tn>t_r\) for the condition used in the main-text numerical tests.
The autocorrelation estimate is obtained at \(k=0\), and we use it as a representative scale for the tested \(k=\nint=3,4,5\) cases, where the stable products are dilute and do not introduce an additional slow variable in the present model. 

\begin{figure}[tb]
\centering
\includegraphics[width=0.5\textwidth]{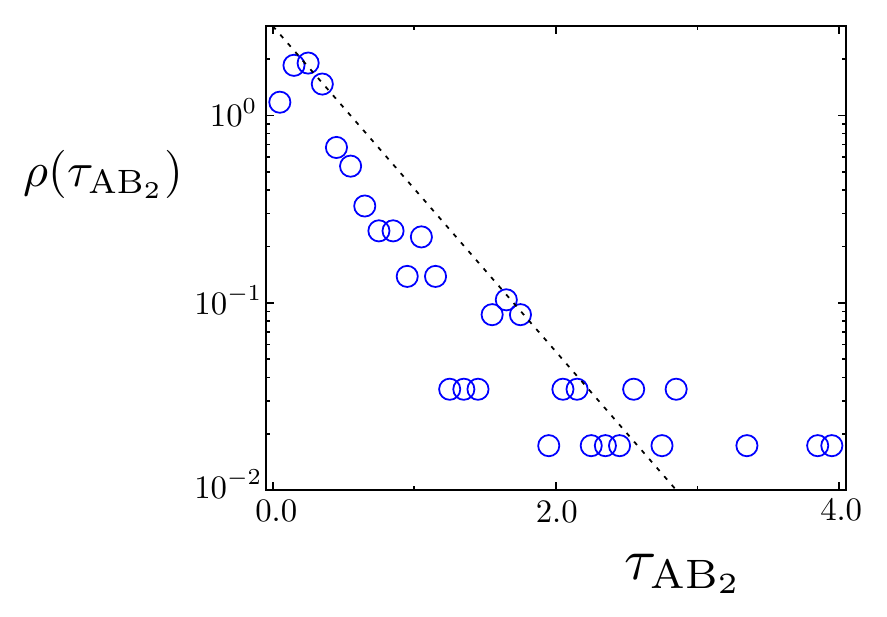}
\caption{
Distribution of raw geometrical \ce{AB2} lifetimes at \(T=2.5\) and \(k=0\).
The dashed line is a guide for the eye proportional to \(\exp(-\tau_{\ce{AB2}}^{\rm raw}/0.5)\), indicating a microscopic contact time \(t_{\rm contact}\simeq0.5\).
}
\label{fig:raw_lifetime_provisional}
\end{figure}

This time-series analysis also resolves fluctuations associated with brief close approaches.
Figure~\ref{fig:raw_lifetime_provisional} shows the distribution of the raw lifetime of a single geometrically identified \ce{AB2} event at \(T=2.5\) and \(k=0\).
This lifetime is characterized by a microscopic time of order \(0.5\),
which we denote by \(t_{\rm contact}\). 
We choose \(\Delta=5\), which is
an order of magnitude longer than \(t_{\rm contact}\).
It is much shorter than the coarse-grained trimer-number time \(\tn\simeq2\times10^2\).
This distinction is important: raw contacts diagnose microscopic recrossings, whereas the time-averaged-distance observable defines the reaction event used for waiting times.

\begin{figure}[tb]
\centering
\includegraphics[width=0.85\textwidth]{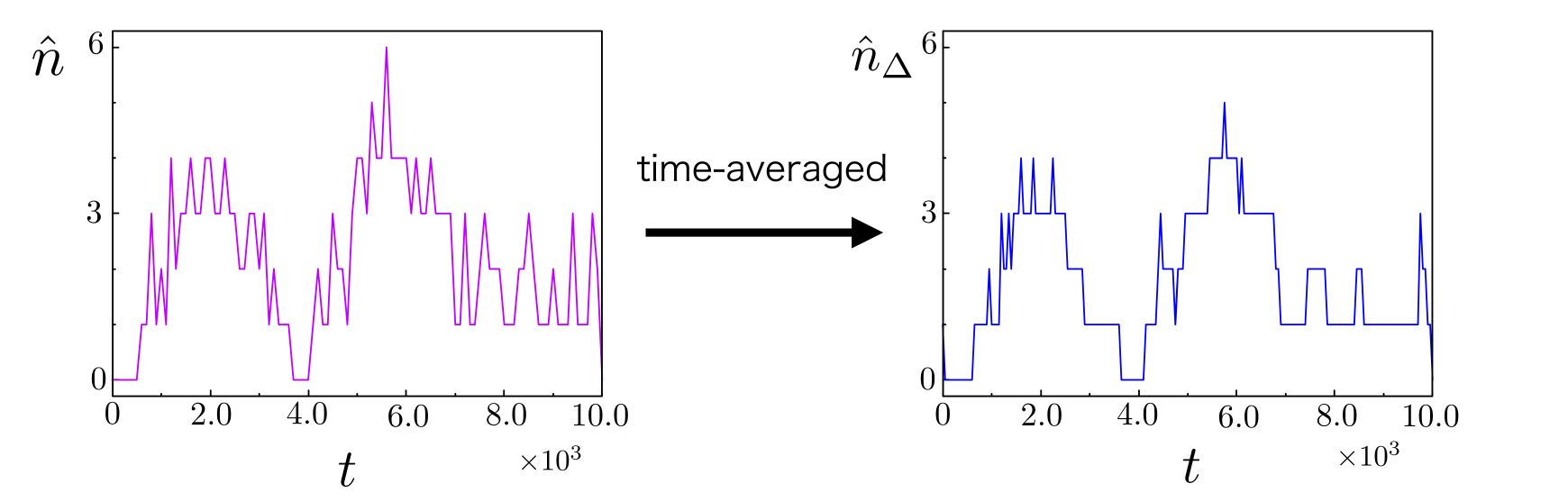}
\caption{
Definition of the coarse-grained trimer number used for the final waiting-time analysis.
The left panel shows the instantaneous geometrical count \(\hat n\), while the right panel shows \(\hat n_\Delta\), obtained by applying the same trimer criterion after averaging the relevant interparticle distances over a window \(\Delta=5.0\).
The time origin is shifted to the beginning of the displayed trajectory window.
This comparison emphasizes that the adopted criterion coarse-grains the molecular geometry itself.
}
\label{fig:resolution_provisional}
\end{figure}

The operational coarse-grained observable used for waiting times is defined at the level of interparticle distances.
Figure~\ref{fig:resolution_provisional} compares the instantaneous geometrical count with the time-averaged-distance count for the same trajectory.

\section{Measurement of waiting times}
\label{s:wtime}

The waiting time \(\tau(n;k)\) is the mean interval from a return to the
typical value \(n_*(k)\) to the next successful arrival at the target value
\(n\), in an equilibrium system with \(k\) pre-existing \ce{C} molecules.
In the single protocol, \(k=0\) and the target is \(n=\nC\).
In the second stage of the staged protocol, \(k=\nint\) and the target is again \(n=\nC\), corresponding to \(\nC-\nint\) newly formed transient complexes in the presence of \(\nint\) stable products.
The second-stage quantity is thus evaluated from equilibrium fluctuations
in the system with \(k=\nint\) stable products.

For the final waiting-time analysis, the successful threshold event is defined by a coarse-grained trimer number \(\hat n_\Delta(t;k)\).
Here \(\hat n_\Delta(t;k)\) is constructed by averaging the relevant interparticle distances over a window \(\Delta=5.0\) and then applying the geometrical trimer criterion defined in Sec.~\hyperref[s:timescale]{VII}.
This is different from simply observing the raw instantaneous trajectory only every \(5.0\) time units, which is not used to define the main-text waiting-time data.
The final Fig.~4 of the main text is constructed for \(\nint=3,4,5\) and \(\nC=\nint+2,\ldots,8\), using \(\tau_1\), \(\tau_{21}\), and \(\tau_{22}\) measured with this same time-averaged-distance definition of \(\hat n_\Delta(t;k)\).

Along an equilibrated trajectory, each interval is measured from the return to the typical value \(n_*(k)\), defined as the most probable value of \(\rho(n|k)\), until the subsequent successful arrival at the target \(n\).
After an event is counted, the trajectory is not counted again until it has returned to \(n_*(k)\).
This procedure reduces repeated counting of the same rare excursion.
For waiting times, the threshold event is counted only after the transient complexes satisfy the persistence criterion described in \hyperref[s:timescale]{Sec.~VII}.
This is the operational reason why the main text refers to mean waiting times rather than to a purely instantaneous mean first-passage time.
The persistence requirement removes microscopic recrossings and affects the attempt time, whereas the rare-event weight in the estimate \(\tau(n;k)\simeq \tau_0/\rho(n|k)\) is governed by the equilibrium tail probability \(\rho(n|k)\).

\begin{figure}[tb]
\centering
\includegraphics[width=0.55\textwidth]{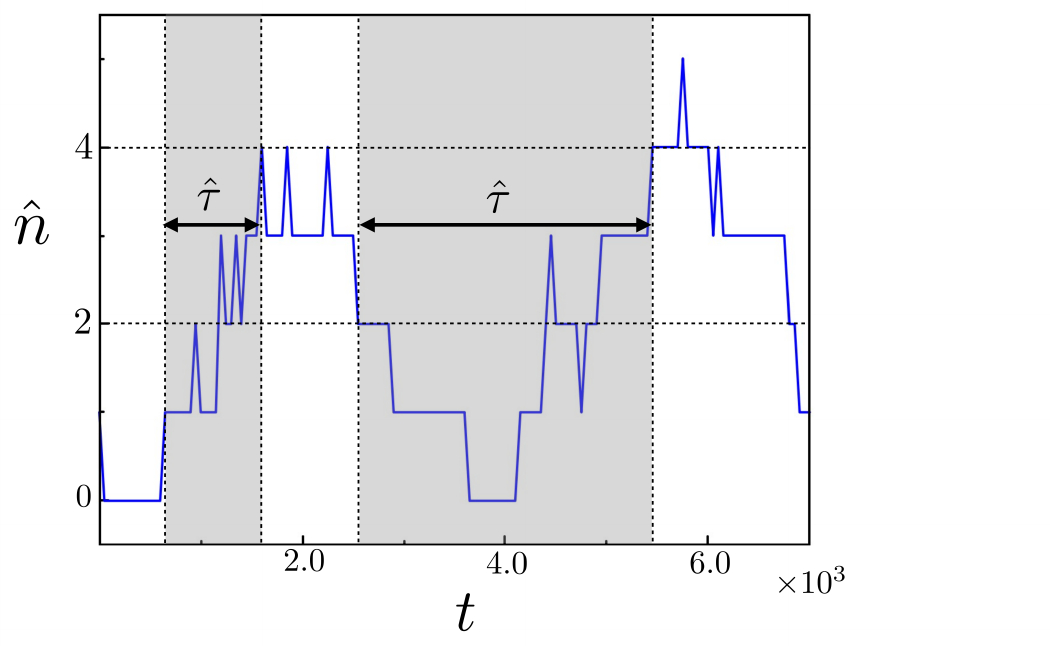}
\caption{
Representative waiting-time intervals measured from the coarse-grained trimer-number trajectory.
The time origin is shifted to the beginning of the displayed trajectory window.
The shaded regions indicate the intervals used to define threshold-event waiting times.
After a threshold event is counted, a new interval is not started until the trajectory returns to the reference value.
}
\label{fig:waiting_event_example}
\end{figure}

Figure~\ref{fig:waiting_event_example} illustrates the operational definition of a waiting-time interval in the coarse-grained trajectory.
In this representative example, the shaded regions indicate intervals measured after the trajectory has returned to the reference level and before the next successful arrival at the target level.
The same return-before-recounting rule is used for the waiting-time data in the main text.

Statistical uncertainties are estimated as standard errors from the threshold-event intervals defined above and are shown as error bars in Figs.~2(b) and 4 of the main text.

This analysis separates two procedures: thinning the raw instantaneous trajectory by a larger observation interval and defining the trimer number from time-averaged interparticle distances.
Only the second procedure is used as the final operational definition of the successful threshold event, because it filters out brief close approaches at the level of the geometrical criterion.
The final analysis plots \(\tau_1/\tau_2\) against the same \(\Ac\) calculated from the instantaneous equilibrium probabilities \(\rho(n|k)\), not from a redefined coarse-grained probability.
The time-averaged-distance rule enters the measured waiting times and hence the non-combinatorial prefactor \(K\).
Over the tested range, the final data for \(\nint=3,4,5\) are described by an empirical common prefactor \(K=0.43\) within standard errors.